\documentclass[journal]{IEEEtran}
\usepackage{textcomp}
\usepackage{cite}
\usepackage{amsmath,amssymb,amsfonts}
\usepackage{algorithmic}
\usepackage{graphicx}
\usepackage{multicol,multirow}
\usepackage{array}
\usepackage{gensymb}
\usepackage{mathtools, cuted}
\usepackage{xcolor}
\usepackage{siunitx}
\usepackage{tikz}

\usepackage{lipsum}

\ifCLASSOPTIONcompsoc
    \usepackage[caption=false, font=normalsize, labelfont=sf, textfont=sf]{subfig}
\else
\usepackage[caption=false, font=footnotesize]{subfig}
\fi

\newcolumntype{L}{>{\raggedright\arraybackslash}p}
\newcolumntype{C}{>{\centering\arraybackslash}m}

\newcommand\copyrighttext{%
  \footnotesize \textcopyright 2020 IEEE. Personal use of this material is permitted.
  Permission from IEEE must be obtained for all other uses, in any current or future
  media, including reprinting/republishing this material for advertising or promotional
  purposes, creating new collective works, for resale or redistribution to servers or
  lists, or reuse of any copyrighted component of this work in other works.
  DOI: 10.1109/TAP.2020.3037800}
\newcommand\copyrightnotice{%
\begin{tikzpicture}[remember picture,overlay]
\node[anchor=south,yshift=3pt] at (current page.south) {\fbox{\parbox{\dimexpr\textwidth-\fboxsep-\fboxrule\relax}{\copyrighttext}}};
\end{tikzpicture}%
}

\begin{document}

\title{Radar Cross Section of Chipless RFID tags and BER Performance}

\author{Michele~Borgese,~\IEEEmembership{Member,~IEEE}, Simone~Genovesi~\IEEEmembership{Senior Member,~IEEE}, Giuliano Manara,~\IEEEmembership{Fellow,~IEEE} and Filippo~Costa,~\IEEEmembership{Senior Member,~IEEE} 

}
\maketitle
\copyrightnotice

\begin{abstract}
\boldmath
The performance of different chipless RFID tag topologies are analysed in terms of Radar Cross Section (RCS) and Bit Error Rate (BER). It is shown that the BER is mainly determined by the tag Radar Cross Section (RCS) once that a standard reading scenario is considered and a fixed size of the tag is chosen.  
It is shown that the arrangement of the resonators in the chipless tag plays a crucial role in determining the cross-polar RCS of the tag. The RCS of the tag is computed theoretically by using array theory where each resonator is treated as a separate scatterer completely characterized by a specific reflection coefficient. Several resonators arrangements (periodic and non-periodic) are compared, keeping the physical area of the tag fixed. Theoretical and experimental analysis demonstrate that the periodic configuration guarantees the maximum achievable RCS thus providing a global lower BER of the chipless RFID communication system. We believe that the BER is the more meaningful and fair figure of merit for comparing the performance of different tags than \SI[detect-weight]{}{bit/cm^2} or \SI[detect-weight]{}{bit/Hz} since the increase of encoded information of the tag is useful only if it can be correctly decoded.  
\end{abstract}


\section{Introduction}
\label{sec:introduction}
The aim of chipless RFID technology is to obtain a radio frequency identification system without the use of an Integrated Circuit (IC) \cite{tedjini2013hold}. The removal of the IC allows achieving numerous advantages such as the reduction of the production costs, the operation in harsh environments and the compatibility with wearable and printable electronics \cite{leenen2009printable}. Similarly to conventional RFID \cite{marrocco2010pervasive}, chipless RFID technology can be employed in scenarios where sensing capabilities are also required \cite{Borgese_chipless_humidity,Vena_chipless_CO2_sensor,lazaro2018chipless,genovesi2018chipless}. 
 
Chipless RFID tags are usually differentiated among those operating in time domain (TD) or in frequency domain (FD) \cite{Karmakar_review}. The most interesting TD tags are based on Surface Acoustic Wave (SAW) technology, which exploits the properties of piezoelectric materials \cite{Karmakar_saw} for electromagnetic/acoustic wave conversion. FD tags seem to be very promising and a large number of operational configurations have been proposed so far. The most employed tags consist of a set of passive resonators arranged in a planar configuration \cite{Vena_chipless_ID, rance2016toward}. Some papers have also proposed some Figures of Merit (FoM) aimed at comparing the performance of different tags \cite{herrojo2019chipless,khan20153, mc2019review, svanda2019platform}. Usually, the FoMs  take into account  the size of the tag and the frequency compression of the peaks but they often neglect the Radar Cross Section (RCS) level of the tag. 

The aim of the present work is to analyse the performance of chipless RFID system from the point of view of a communication system \cite{carlson2010communication,chen2019information}. The most relevant parameter to evaluate the performance of the system is not the physical area occupied by the transponder or the bandwidth for encoding information but the capability of the entire system to maintain a certain level of reliability in a standardized measurement scenario. The parameter commonly used for this purpose is the Bit Error Rate (BER) \cite{jeruchim1984techniques}. 
In this view, the reduction of the size of the tag may be not a good choice if the BER increases. Indeed, if the tag is shrunk, its RCS decreases with a consequent reduction of the back-scattered power and thus of the detection probability. As a consequence, the choice of a tag with the smallest size is not necessarily the best choice. The miniaturization of the tag is desirable only if the detection probability is maintained above an acceptable level in a typical operative scenario.
A possible configuration to increase the RCS level of the tag consists in replicating a multi-frequency resonator along one or two planar directions so as to form a periodic surface \cite{Costa_chipless_2013}. This replication does not increase the information contained in the tag, but it greatly improves the detection probability because of the increase of the RCS level. To substantiate this, in this work the RCS of different tag configurations is evaluated both theoretically and experimentally and the disposition of the resonators to synthesize a chipless RFID tag is addressed. In particular, the periodic arrangement of multi-frequency resonators is compared with non-periodic placement of the same resonators in the same physical area. 
It is shown that the former approach leads to a higher mutual interaction among the resonators but it allows achieving a much better RCS level with respect to the aperiodic disposition (as high as \SI{15}{\decibel}). In fact, all the scatterers comprised in the periodic tag reflect the impinging field with the same amplitude and phase whereas an aperiodic disposition of resonators determines an incoherent reflection since each element responds with different amplitude and phase \cite{huang2005reflectarray,Boggese_Reflectarray}.
In addition to that, keeping in mind both array and FSS theory \cite{munk2000frequency,costametamaterials}, also the element spacing has to be adequately considered to avoid the onset of grating lobes.

The paper is organized as follows. Section II describes the  chipless RFID system model employed to define the BER. Section III illustrates a fast semi-analytic methodology for computing the RCS of chipless RFID tags. In Section IV, three different tag configurations are analyzed from the point of view of cross-polar RCS. Section V reports the experimental results achieved with two different tag configurations. Finally, Section VI is dedicated to comparing the performance of the different tag configurations in terms of BER. Concluding remarks are reported in Section VII.   

\section{Chipless RFID System}  
\begin{figure}
\centering
\centerline{\includegraphics[width=8cm,keepaspectratio]
{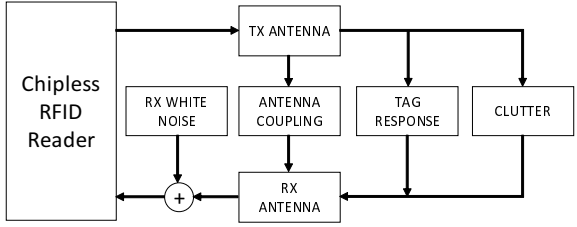}}
\caption{Schematic representation of the chipless RFID communication system with the main signal contributions collected at the receiver.}
\label{fig:TxRx_scheme}
\end{figure}

A chipless RFID can be seen as a radar communication system. The main parameter to evaluate its performance is the Probability of Error or Bit Error Rate (BER) \cite{carlson2010communication, Costa_ISAR}. Commonly used figures of merit, such as \SI[detect-weight]{}{bit/cm^2} or \SI[detect-weight]{}{bit/Hz}, do not reveal that the increase of the information encoded in a tag is useful only if it can be correctly decoded. The BER can be evaluated by modelling the received signal as the sum of the reflected transmitted signal and detrimental contributions represented by the antenna coupling, the clutter and the receiver noise. The contributions which determine the shape and the level of the received signal in a chipless system are summarized in Fig.~\ref{fig:TxRx_scheme}. The useful contribution which contains the information provided by the tag is often weak and can be overwhelmed by the undesired contributions leading to the impossibility of detecting the bit sequence or the information encoded in the tag. 

The intensity of the useful signal, $S^{tag}$, mainly depends on the RCS of the target, $RCS^{tag}$, which largely varies for the different tag configurations. In particular, the received power can be computed according to the classical radar equation:

\begin{equation} \label{eq:PR} 
S^{tag} = \frac{G_R^2 P_T RCS^{tag} \lambda^2}{(4\pi)^3 d^4} 
\end{equation}

where $\lambda$ represents the wavelength, $d$ represents the distance from the reader antenna to the tag, $G_R$ is the gain of the reader antenna and $P_T$ is the transmitted power. Keeping unchanged the measurement setup (distance between the tag and the reader, the gain of the reader antenna and the transmitted power), the received power depends only on the RCS of the tag and on the operating wavelength. The RCS of a scattering tag can be approximated as the product of the RCS of the metallic square plate occupying the same area of the tag ($RCS^{plate}$) and the reflection coefficient of the tag as:

\begin{equation} \label{eq:RCS_tag} 
RCS^{tag} = RCS^{plate} (\Gamma^{tag})^2=\frac{4 \pi A^{tag}}{ \lambda^2} (\Gamma^{tag})^2
\end{equation}

The unwanted signal at the receiver sums up to the useful contribution reflected by the tag and deteriorates the system performance. The bit sequence embedded within the tag is correctly detected if the signal level emerges from the noise level. According to the scheme reported in Fig.~\ref{fig:TxRx_scheme}, the received signal can be modelled as:

\begin{equation} \label{eq:rec_signal} 
S^{RX} = S^{tag} + S^{coupling} + S^{clutter} + S^{noise}
\end{equation}
 
In a radar system \cite{skolnik1980introduction}, the clutter contribution usually exceeds the random noise at the receiver. Assuming that the coupling contribution can be largely removed by using time domain gating \cite{Ramos_Temporal, Costa_Normalization} or by making a background subtraction \cite{Costa_chipless_2013} (the backscattered signal from the environment arrives after the coupling signal in time), the received signal can be roughly approximated as:

\begin{equation} \label{eq:rec_signal_approx} 
S^{RX} \simeq S^{tag} + S^{clutter}
\end{equation}

In this simplified scenario, it is clear that the detection of the bit sequence is possible if the useful signal exceeds the clutter contribution. Assuming a standard indoor scenario, the clutter can be modeled as a complex normal random process:

\begin{equation} \label{eq:clutter_model} 
S^{clutter} = \mathcal{CN}(\mu,\sigma^2) = C_R + C_I
\end{equation}

The variables $C_R$ and $C_I$ are Gaussian processes with mean $\mu$ and standard deviation $\sigma$. The amplitude of the complex normal random process is characterized by a Rice distribution and a phase is characterized by a uniform distribution with values between $ \pi $  and $ -\pi $. The parameters used to model the amplitude and the phase of random processes are summarized in Table \ref{tab:noise_signal}. A typical value of the clutter amplitude in a complex indoor environment is $-52.6$~dB for the amplitude and $2.3$~dB for the standard deviation \cite{d2012indoor}.

\begin{table}[ht]
\caption{Parameters used to model clutter.}
\centering
\begin{tabular}{|l|c|c|c|}
\hline
 & \textbf{Process Type} & \textbf{Average value} & \textbf{Standard Deviation} \\
\hline
$C_R$ & Gaussian & -53 dB & 2.3 dB \\
\hline
$C_I$ & Gaussian & -53 dB & 2.3 dB \\
\hline
\end{tabular}
\label{tab:noise_signal}
\end{table}

In summary, when we deal with a chipless RFID system, it is important to know that the received signal intensity has to emerge from a noise floor in order to achieve the correct bit sequence detection. Therefore, in order to perform a correct detection, it is important to have a high RCS of the tag and not only intelligible peaks (waveform). 

\section{RCS of a Chipless Tag} 
\label{sec:RCS}
\begin{figure}
    \centering
  \subfloat[]{%
       \includegraphics[width=4.4cm,height=4.4cm,keepaspectratio]{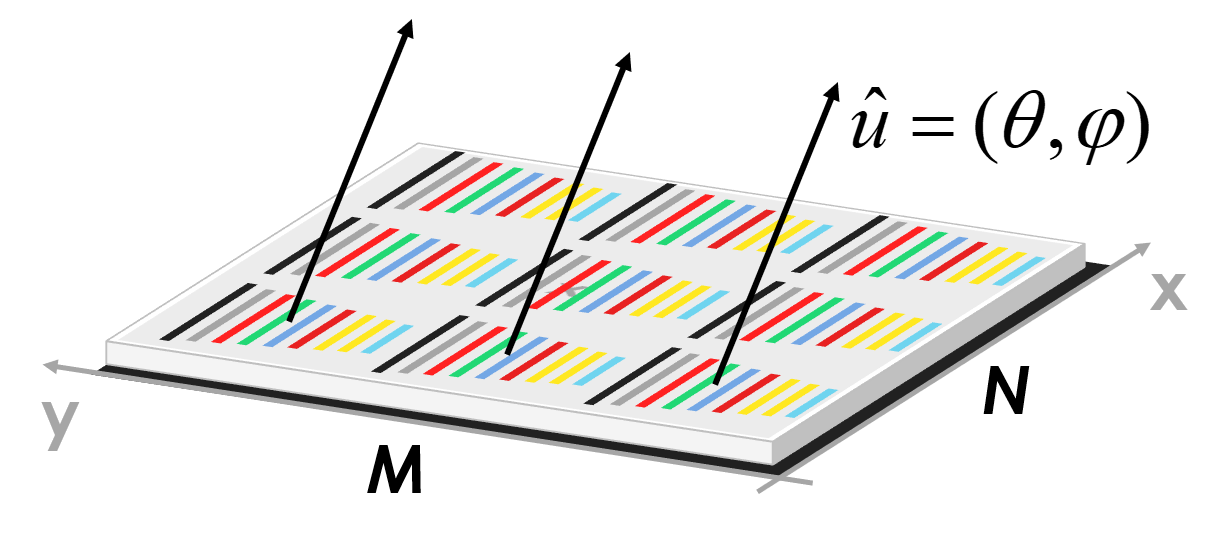}}
    \hfill
  \subfloat[]{%
        \includegraphics[width=4.4cm,height=4.4cm,keepaspectratio]{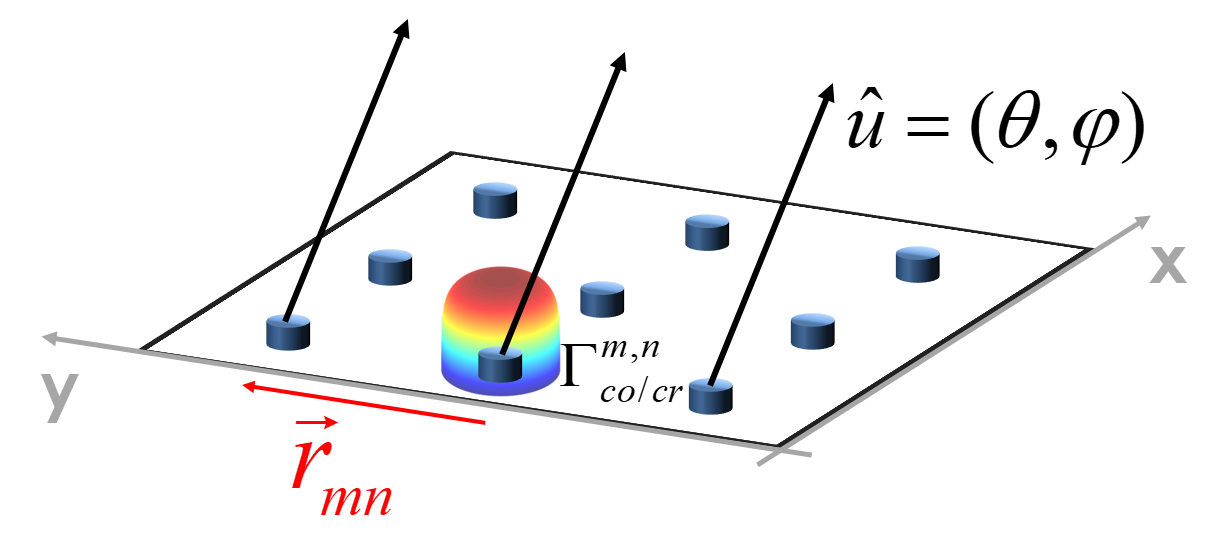}}     
      \caption{(a) $ (M\times N) $ - chipless tag in the $ (x,y) $ coordinate system; (b) discrete scatterers model of a generic chipless tag.  
      \label{fig:Geometry}}      
\end{figure}

The RCS of a chipless tag can be computed by using full-wave solvers but it may require a considerable amount of time since the resonators are characterized by a high-quality factor and the simulations do not rapidly converge. In addition to that, full-wave solvers do not provide any physical insight into the scattering mechanisms.
A more efficient and insightful determination of finite-size chipless tag scattering relies on planar array and reflectarray theory \cite{stutzman, Boggese_Reflectarray} together with physical optics (PO) theory \cite{ruck} and Periodic Method of Moments (PMM). 
According to the Physical Optics (PO) the RCS of a metallic plate can be computed with closed form equations which depend on the geometry of the plate. As is well known, an infinite extent metallic plate provides a unitary reflection, thus its reflection coefficient is equal to -1. If the metallic plate is covered with some absorbing or polarization sensitive material capable of altering the reflection coefficient, the RCS of the coated plate can be computed by weighting the RCS level of the metallic plate with the value of the reflection coefficient of the infinite extent surface according to \cite{balanis1999advanced,catedra2008new}. PMM is used to compute the reflection coefficient of scattering particles and  reflectarray theory is employed to determine the total scattering due to different scattering particles.
In order to illustrate this method, a generic chipless tag in the $ (x,y) $ coordinate system is considered (Fig.~\ref{fig:Geometry}~(a)). The tag can be modelled with $ M \times N $ discrete scatters spaced of $ D = \left|{\vec r}_{mn}\right|$ each one characterized by its copolar/crosspolar reflection coefficient  $ \Gamma^{m,n}_{co/cr} $ and its radiation pattern represented with a $ \cos^{q_e}\left( \theta \right) $ function as shown in Fig.~\ref{fig:Geometry}~(b).

The scattering intensity of a chipless tag towards a generic direction $ \hat{u}$  is directly proportional to the physical area of the scattering object and it depends on the interference among all the point scattering contributions identified on the object. Since the chipless tag can be seen as a summation of complex contributions given by different unit cells (each of them characterized by a specific complex reflection coefficient), the global scattering pattern at the operating frequency $f$ can be computed similarly to an antenna array as: 

\begin{multline}
\vec E^r_{co/cr}(\hat u,f) = \sum\limits_{m = 1}^M {\sum\limits_{n = 1}^N {{{\cos }^{{q_e}}}(\theta )} } \,| {\Gamma^{m,n}_{co/cr} (\hat u,f)}|\\ 
{e^{ - jk({{\vec r}_{mn}} \cdot \,\hat u)}} {e^{ - j\left( {\angle{\Gamma^{m,n}_{co/cr} (\hat u,f)}} \right)}}
\end{multline} 

The argument of the exponential function takes into account the phase shift that is experienced by the wave along the specific pointing direction $ \hat u $. The terms $ |{\Gamma^{m,n}_{co/cr} (\hat u,f)}| $ and $ {e^{ - j\left( {\angle{\Gamma^{m,n}_{co/cr} (\hat u,f)}} \right)}} $ represent the reflection coefficient of the $(m,n)-th$ cell, which is evaluated by a Periodic method of Moment (PMM) code \cite{mittraFSS}. The PMM is a full-wave solver that provides an accurate estimation of the reflection coefficient of an infinite periodic surface embedded within multilayer media by using the Floquet Theorem. The PMM solver is a dedicated approach for solving periodic structures and it is therefore extremely fast. However, this approach cannot be used to compute the RCS of a finite structure as it is specifically dedicated to infinite problems.  The RCS of the tag is computed by weighting the normalized pattern towards the chosen direction with the RCS of a plate of the same area of the total scatterer as follows: 

\begin{equation} \label{eq:RCS} 
RCS^{tag}_{co/cr}(\hat u,f) = \frac{4\pi A^2}{\lambda^2} \frac{|\vec E^r_{co/cr}(\hat u,f)|}{max\lbrace | \vec E^r_{co/cr}(\hat u,f)| \rbrace} 
\end{equation}

where $ A $ is the geometric area of the chipless tag and $ \lambda $ is the operating wavelength. 
By employing the proposed approach, the RCS of the tag can be computed towards a generic direction but the monostatic contribution is the one of interest in case of chipless RFID application. Once the direction to be analysed is chosen, the RCS can be computed within the frequency bandwidth of interest. 
The RCS of finite structures can be computed also with commercial solvers based on Integral Equation (Feko), time or frequency domain (CST) or FEM (HFSS). All the three approaches require extremely long computation times and often the results are not enough accurate. The main hurdle in computing the RCS of these finite structures hosting several resonators placed close to a ground plane is that the resonators  and the metallic surface act as a Fabry-Perot interference device with multiple reflection contributions involved. In order to accurately capture these phenomena a very fine local mesh is required to achieve good results.  The most accurate and stable results have been obtained by using Ansys HFSS and the computation time is in the order of 25 hours  with some variations depending on the mesh accuracy, solution frequency, computational resources and on the specific geometry of the structure. The proposed approach can computed the RCS of these resonant structures in the order of a few tents of seconds.

\section{Numerical results} 
The effectiveness of the scatterer disposition in the design of chipless RFID tags can be analysed by using the proposed formulation. The maximum RCS obtainable with a tag characterized by a given physical dimension is represented by the physical optics relation in eq.~(\ref{eq:RCS}). 

\begin{figure}
    \centering
  \subfloat[][\scalebox{0.9}{\textit{Periodic}}]{%
       \includegraphics[width=2cm,keepaspectratio]{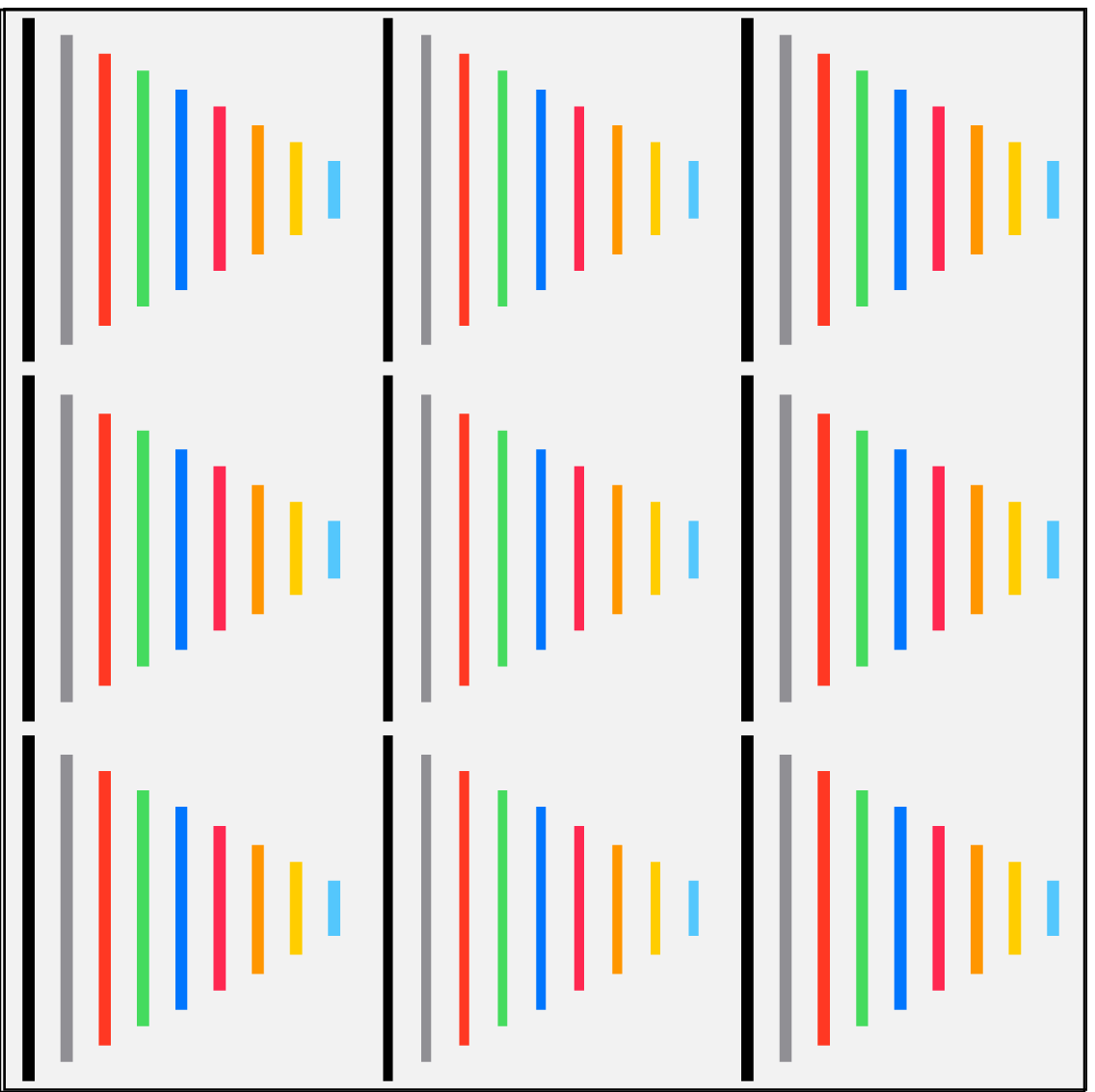}}
    \label{}\hfill        
  \subfloat[][\textit{\scalebox{0.9}{\textit{NonPeriodic-9}}}]{%
        \includegraphics[width=2cm,keepaspectratio]{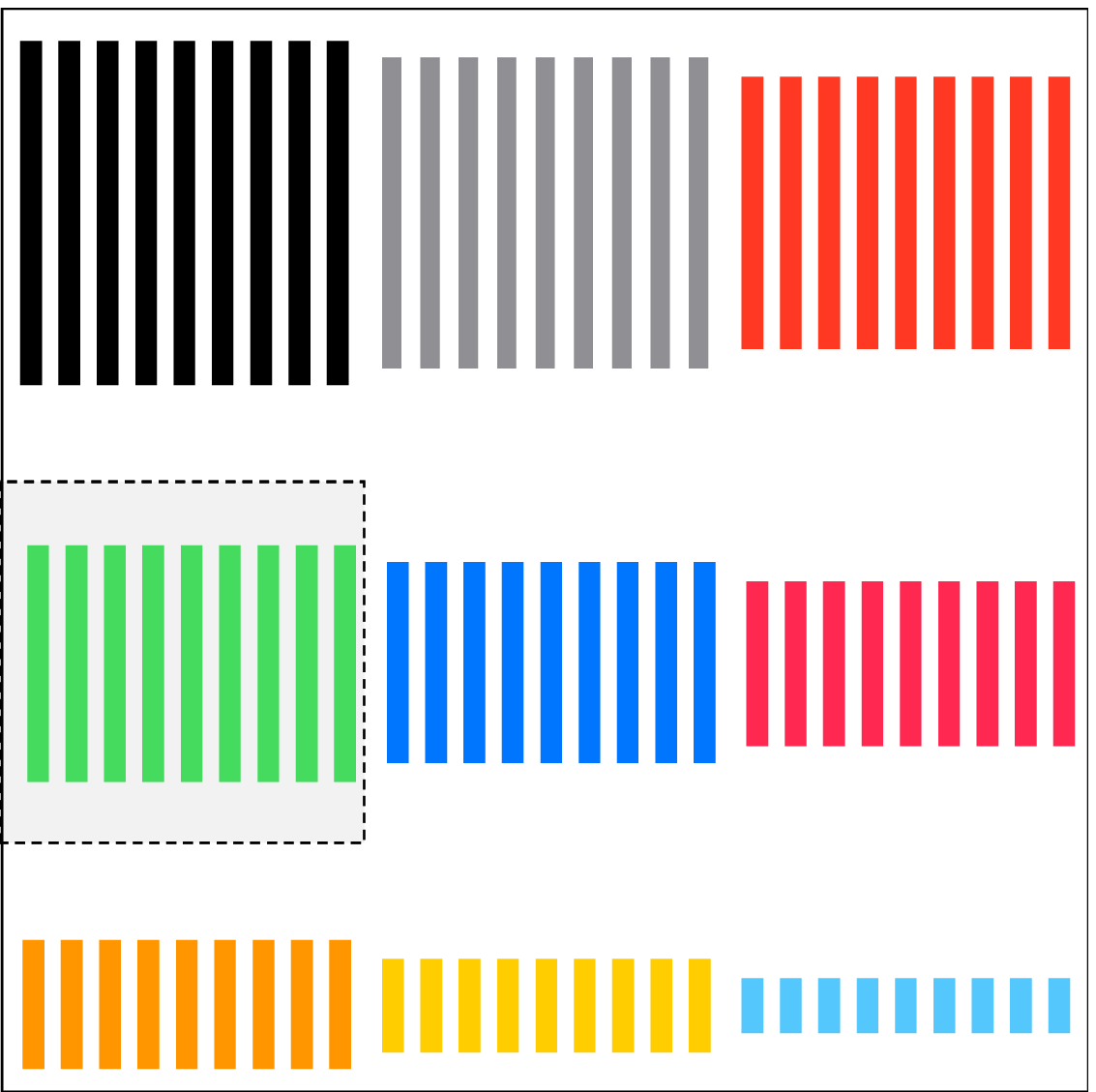}}
\label{}\hfill        
  \subfloat[][\textit{\scalebox{0.9}{\textit{NonPeriodic-5}}}]{%
        \includegraphics[width=2cm,keepaspectratio]{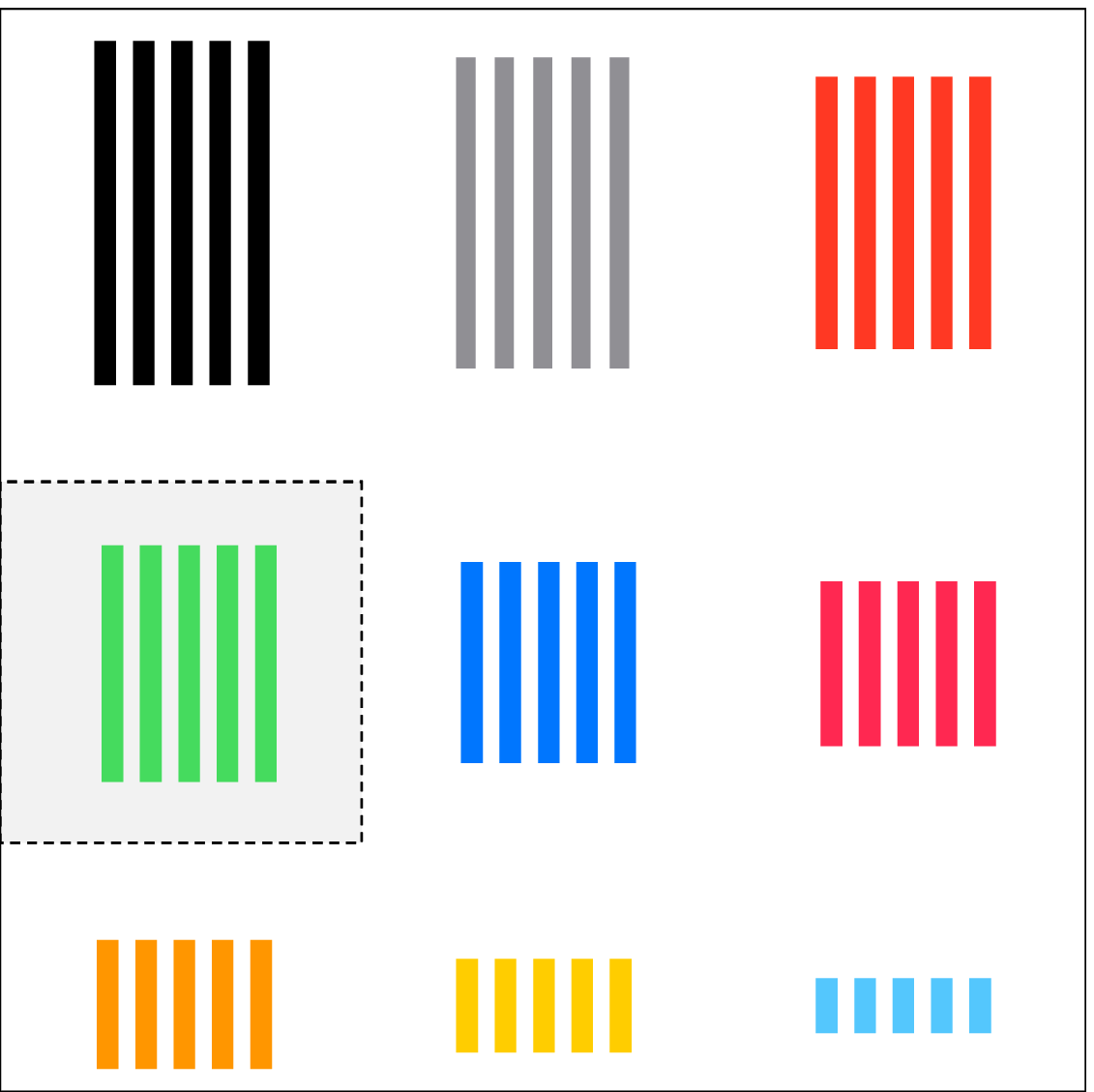}}
    \label{}\hfill
  \subfloat[][\scalebox{0.9}{\textit{NonPeriodic-1}}]{%
        \includegraphics[width=2cm,keepaspectratio]{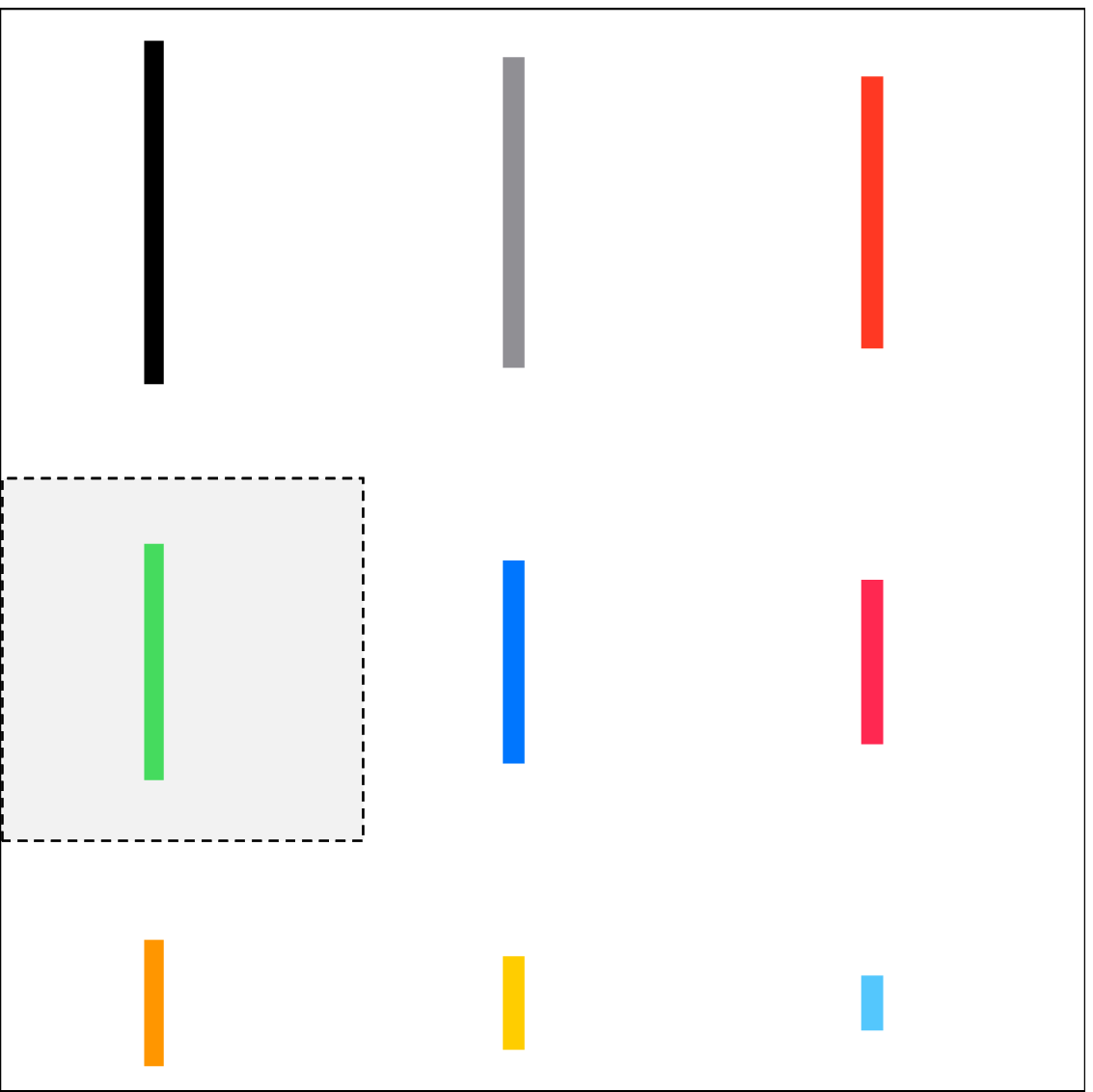}}
\label{}\hfill
  \caption{Four different configurations of  dipole-based chipless tags characterized by the same physical area. (a) 81 $(9\times9)$ dipoles arranged in a periodic configuration;(b) 81 $(9\times9)$ dipoles,  (c) 45 $(5\times9)$ dipoles and (d) 9 $(1\times9)$  dipoles arranged in a non-periodic configuration. The area of the tag which provides an in-phase response is highlighted in grey.}
  \label{fig:in_phase_area} 
\end{figure}

\begin{figure} 
    \centering
  \subfloat[][This method]{%
       \includegraphics[width=4.3cm,keepaspectratio]{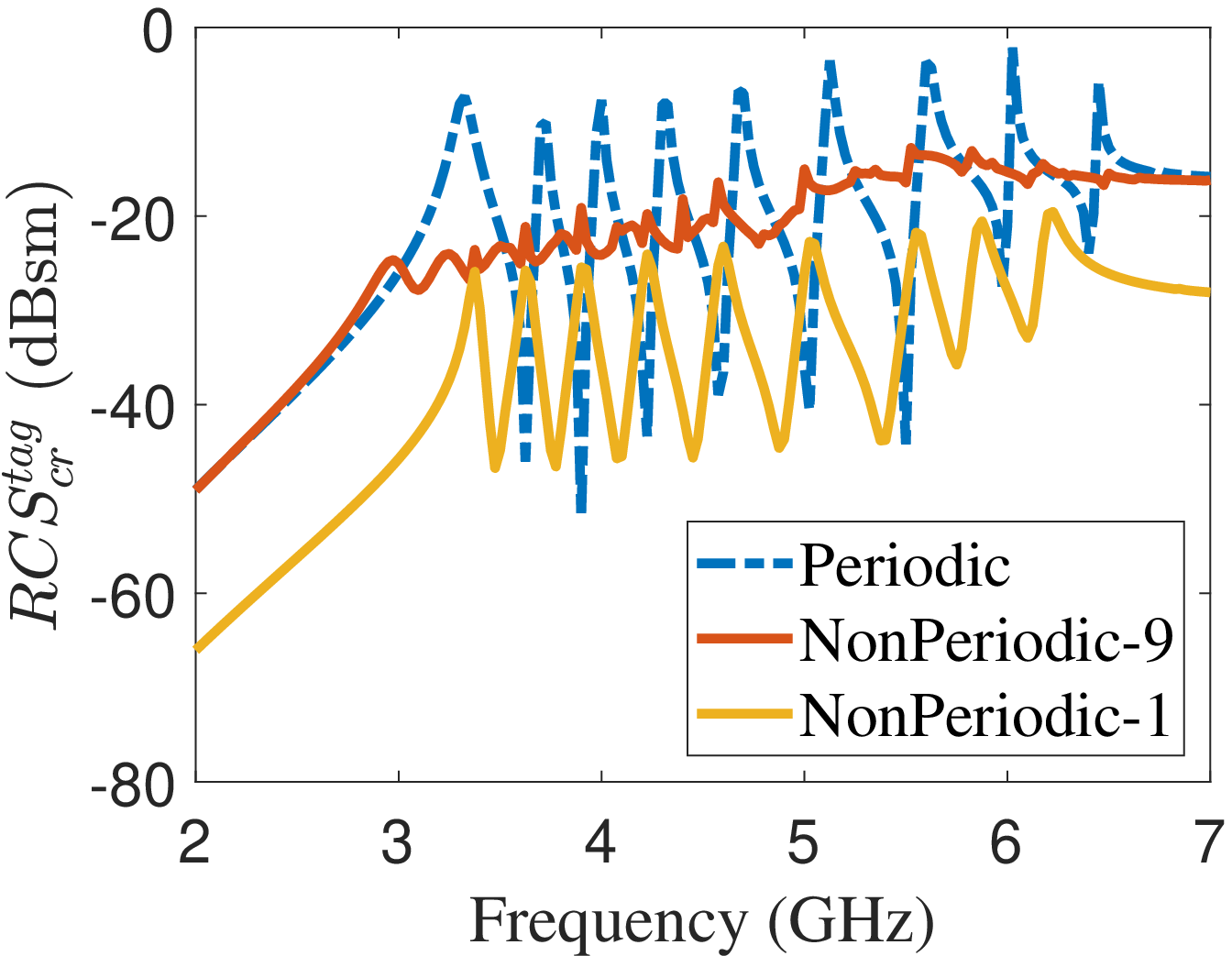}}
     \label{}\hfill
  \subfloat[][HFSS]{%
        \includegraphics[width=4.3cm,keepaspectratio]{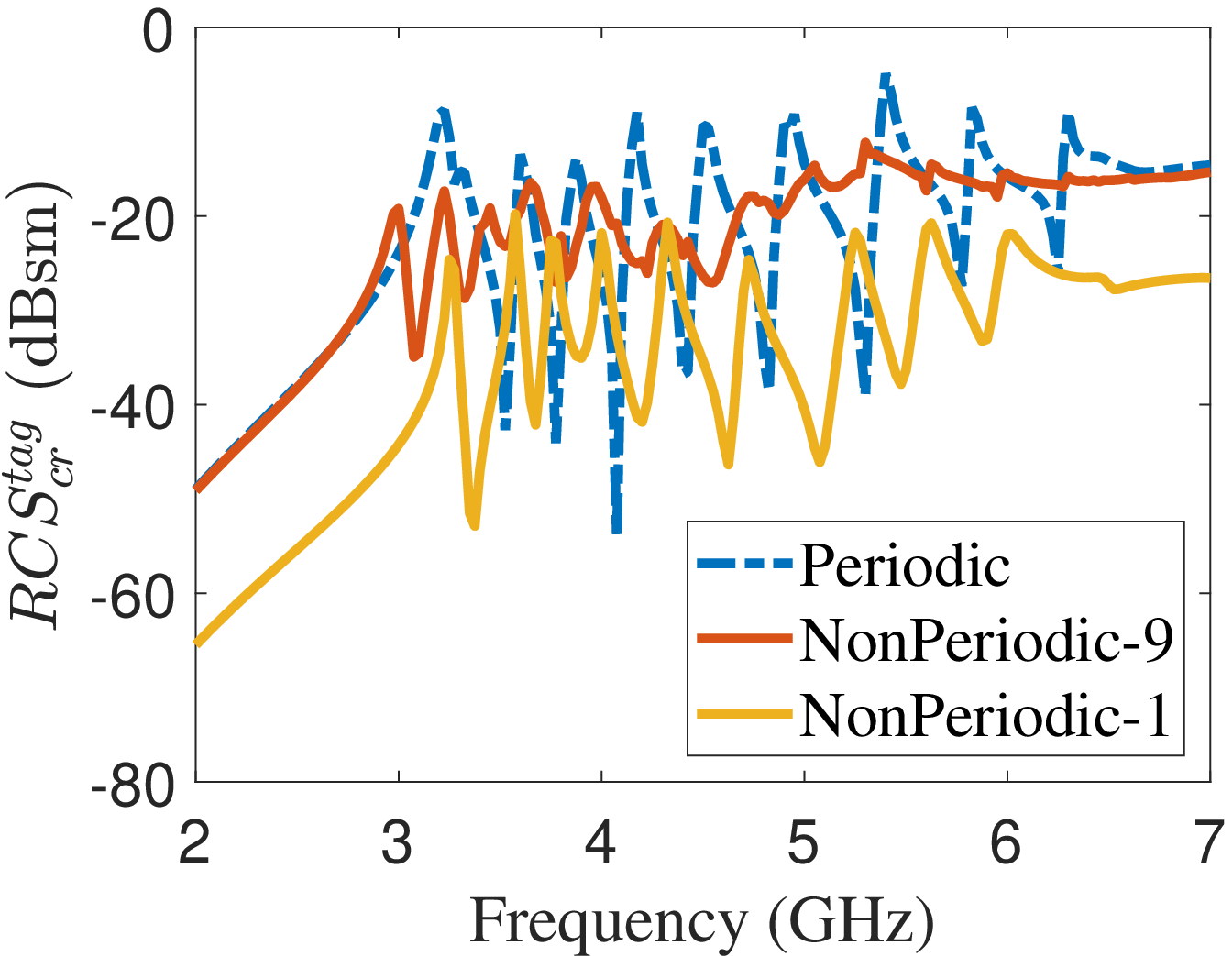}}
         \label{}\hfill
          \caption{Crosspolar RCS in the case of a dipole-based chipless tag arranged in a periodic and non-periodic configurations (NonPeriodic-9 and NonPeriodic-1). (a) Proposed approach, (b) Ansys HFSS. The substrate is 2 mm thick and it is characterized by a dielectric permittivity equal to 2.08.}
      \label{fig:RCS_dipoles} 
\end{figure}

\begin{figure} 
    \centering
  \subfloat[][\textit{Periodic}]{%
       \includegraphics[width=4.3cm,keepaspectratio]{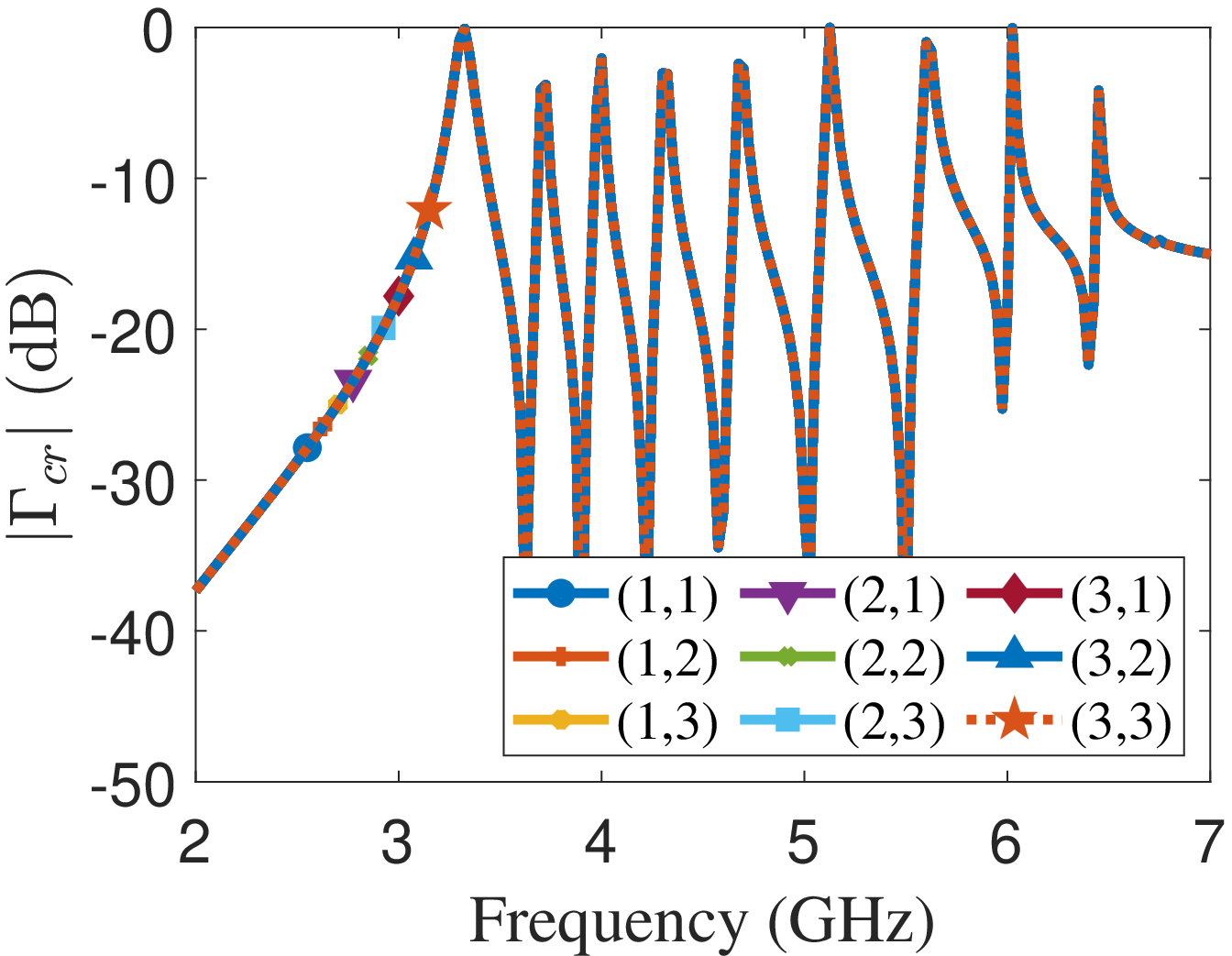}}
     \label{}\hfill
  \subfloat[][\textit{NonPeriodic-9}]{%
        \includegraphics[width=4.3cm,keepaspectratio]{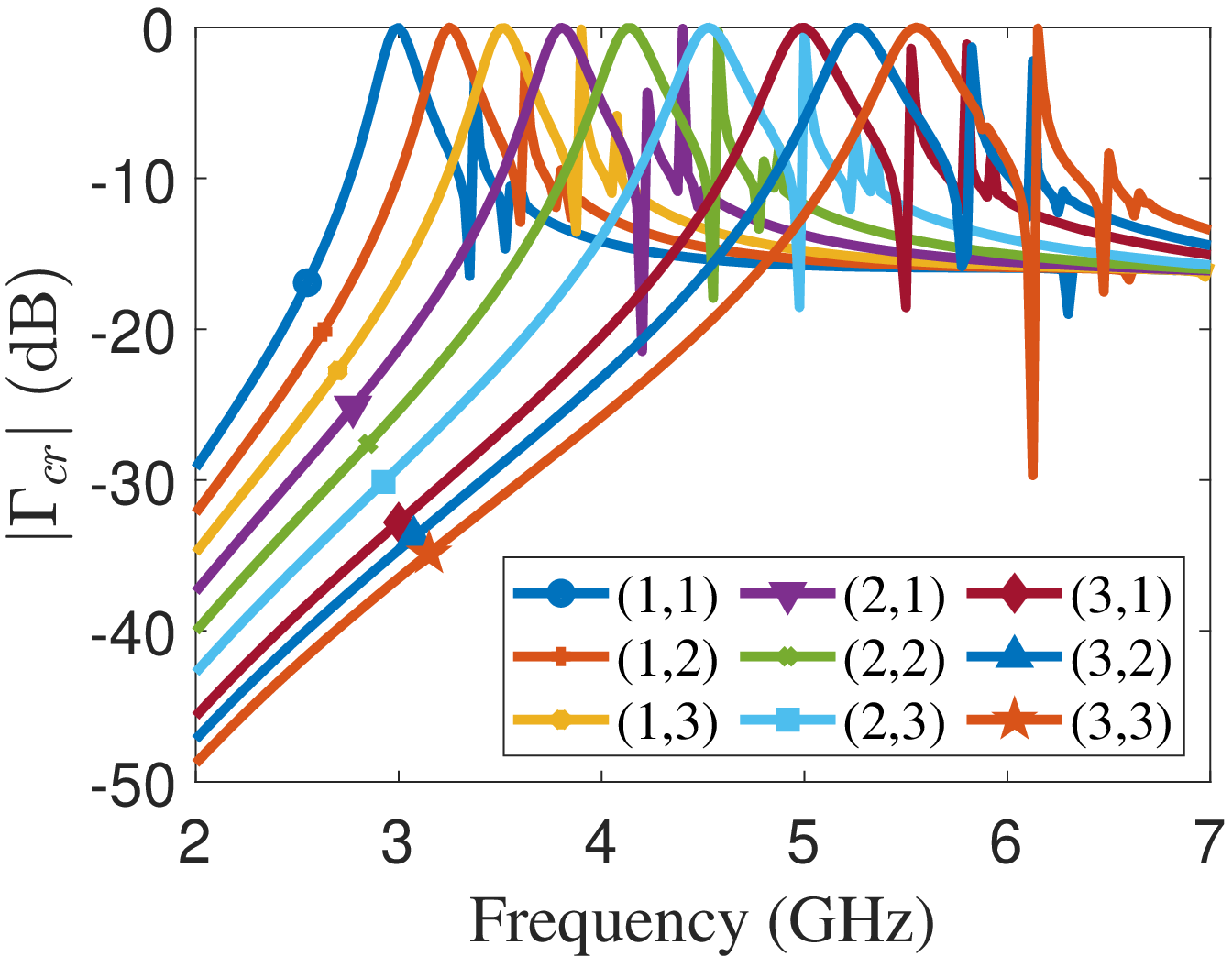}}
         \label{}\hfill
      \subfloat[][\textit{NonPeriodic-5}]{%
        \includegraphics[width=4.3cm,keepaspectratio]{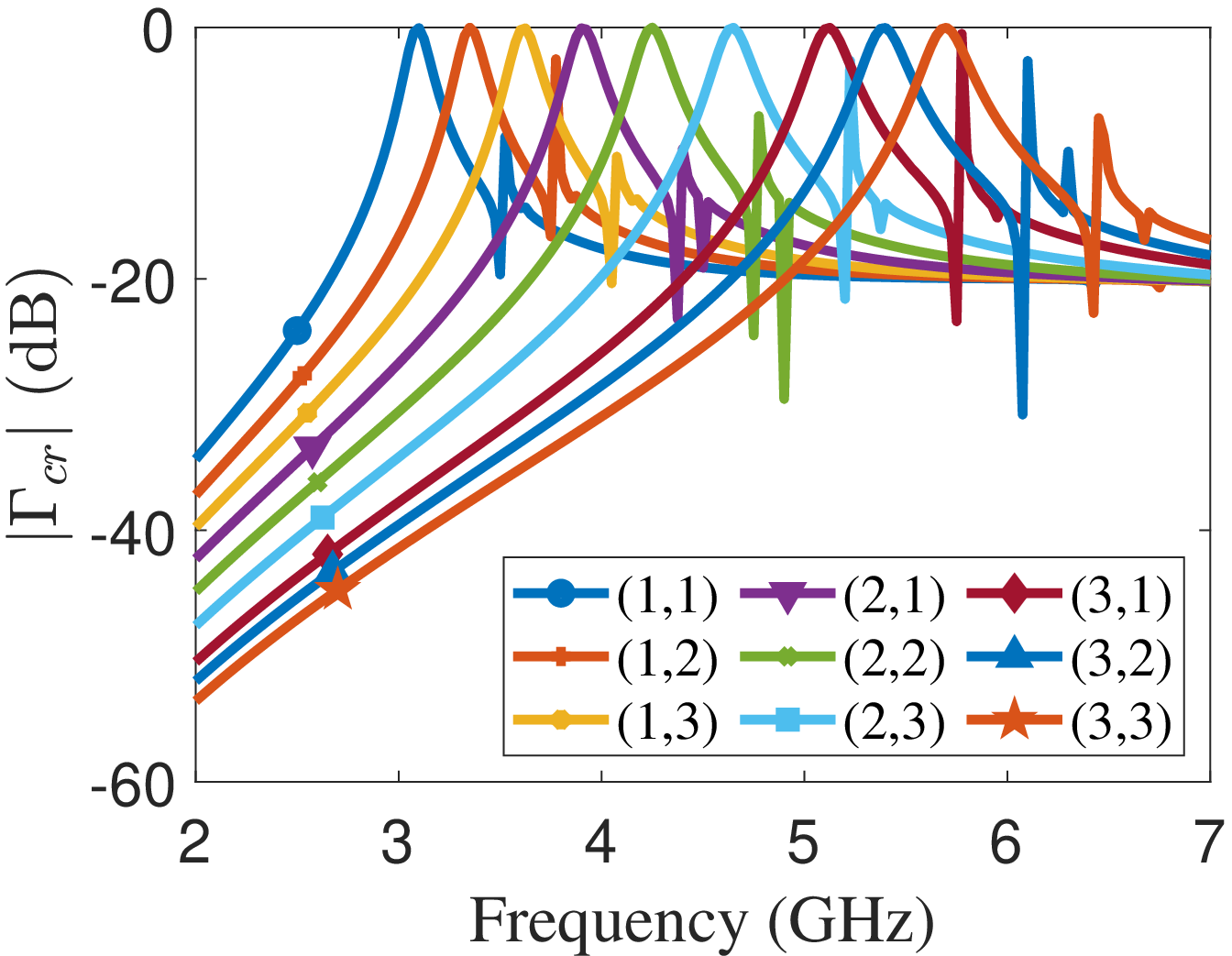}}
\label{}\hfill
      \subfloat[][\textit{NonPeriodic-1}]{%
        \includegraphics[width=4.3cm,keepaspectratio]{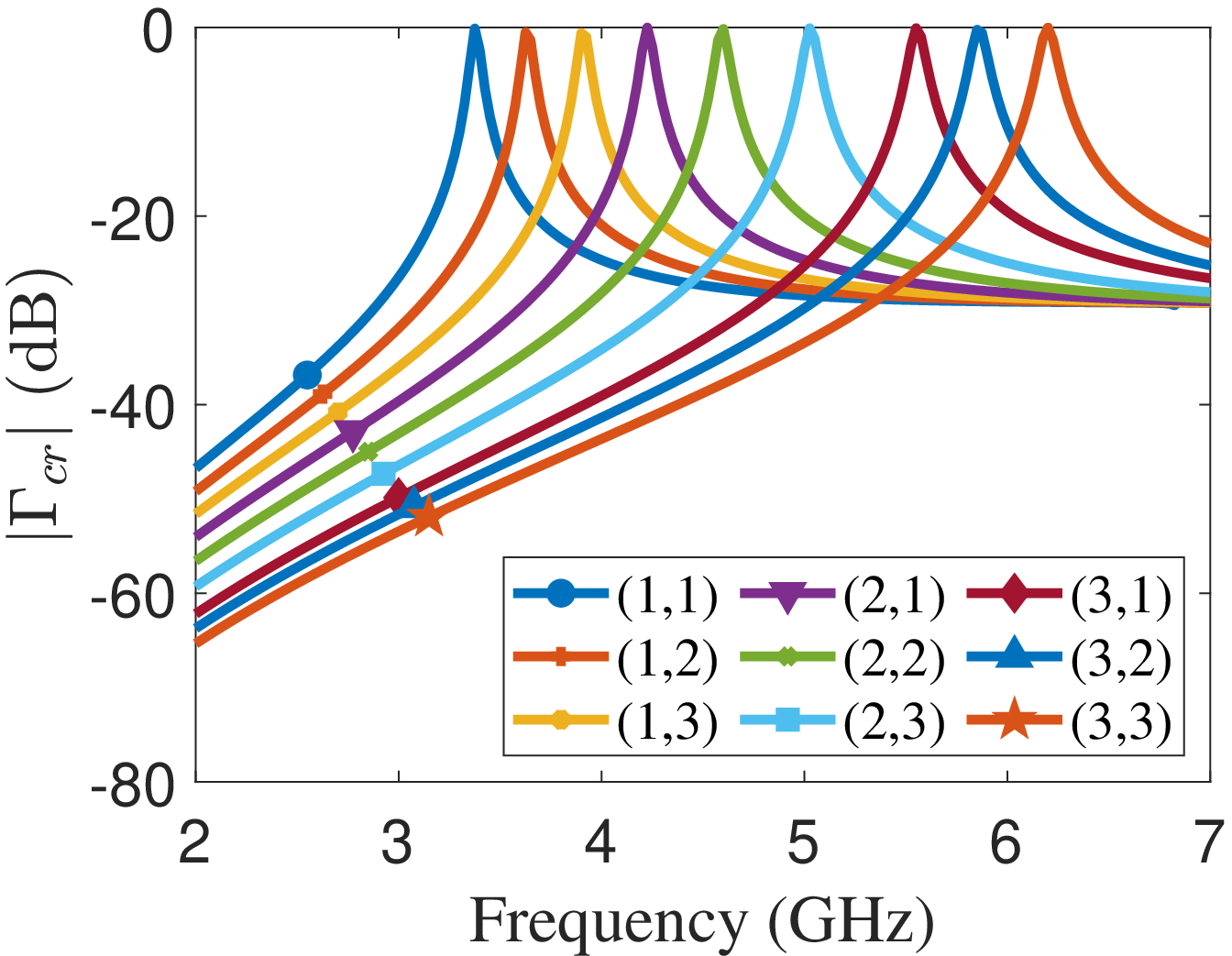}}      
      \caption{Cross-polar reflection coefficient of all the (m,n)-th cells: (a) \textit{Periodic}, (b) \textit{NonPeriodic-9}, (c) \textit{NonPeriodic-5}, (d) \textit{NonPeriodic-1}.}
      \label{fig:amplitude_reflection_coefficients} 
\end{figure}

In case of normal incidence, according to eq.~(\ref{eq:RCS}), this limit can be reached if all point sources are characterized by the same reflection coefficient (amplitude equal to $1$ with all the point sources in phase). In case of off-normal incidence, the monostatic scattering could be maximized, for a specific angular direction, by selecting a specific scattering gradient in the phase reflection coefficient of the point sources according to the theory of reflectarray antennas \cite{Boggese_Reflectarray}. However, the aforementioned case is not specifically interesting for chipless tags comprising a limited number of resonators as the scattered field is not selective in space. 
The four different tag configurations, shown in Fig.~\ref{fig:in_phase_area}, are considered for comparison. Dipole resonators are selected as frequency tuned scatterers \cite{Vena_depolarizing}. The dipole chipless tag can work with co-polar component \cite{Costa_chipless_2013, Manteghi_chipless} but, more interestingly, is able to perfectly convert the impinging polarization into the cross-polar one at a specific frequency \cite{Vena_depolarizing}. In order to maximize  the cross-polar scattering and thus obtaining a perfect polarization conversion, it is necessary to select a surface able to provide a reflection coefficient equal to one in amplitude for both vertical and horizontal polarization and a reflection phase difference of 180$^{\circ}$ between vertical and horizontal polarization \cite{genovesi_orientation}. An alternative suitable resonator for this purpose is a rectangular loop \cite{Costa_differential_encoding}. A detailed explanation of the polarization conversion mechanism is provided in appendix \ref{Appendix-A}. 
In the case of dipole resonators \cite{borgese_costa_circuit}, the scatterers can be disposed in a compact space so as to form a multi-frequency unit cell of a periodic surface or it can be arranged with a larger spacing in a grid so as to form an aperiodic arrangement. In this latter case, the dipoles can be repeated (e.g. \textit{NonPeriodic-9}, \textit{NonPeriodic-5} ) or isolated (\textit{NonPeriodic-1}) \cite{Vena_depolarizing}. The arrangement of the dipoles at a certain distance provides a good immunity to mutual coupling. Moreover, the localized disposition of the resonators can be also used to associate a specific frequency peak to a specific zone of the tags for gesture sensing applications \cite{barbot2017gesture}. However, the aperiodic configuration is prone to a reduction of the global cross-polar RCS due to interference among different scatterers. 
The analysed tags can be seen as a 9 points source scatterers, where each point source is characterized by a reflection coefficient with specific amplitude and phase. As previously mentioned, at normal incidence, the monostatic RCS is maximized if the 9 points sources are in phase and are characterized by a reflection coefficient equal to one in amplitude. When the $9$ resonant scatterers are contained in a single cell and each unit cell is replicated $9$ times to occupy the desired area, all the scattering centres are characterized by the same amplitude and phase. On the contrary, if the nine dipoles are arranged in a non-periodic fashion, the nine point sources have a different amplitude and phase reflection coefficient. In the latter case, the monostatic RCS is, by definition, lower than the maximum achievable one according to physical optics. 
The cross-polar RCS of the three investigated configurations are reported in Fig.~\ref{fig:RCS_dipoles}. 

\begin{figure}
    \centering
       \includegraphics[width=7cm,keepaspectratio]{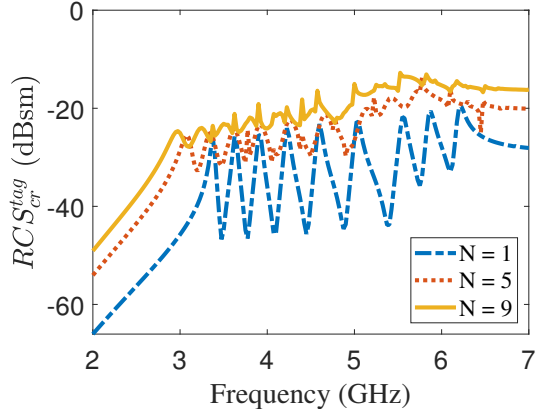}
        \caption{Effect of the number of dipoles $N$ in the NonPeriodic configurations: $N=9$ (\textit{NonPeriodic-9}), $N=5$ (\textit{NonPeriodic-5}), $N=1$ (\textit{NonPeriodic-1}). }
      \label{fig:RCS_vs_Ndipoles} 
\end{figure}

The RCS has been computed both by the proposed semi-analytical formulation and by using a full-wave solver (Ansys HFSS). The general behaviour of the different types of arrangements are in agreement and thus the validity of the proposed RCS computation approach is confirmed. By observing the behaviour of the curves, it is evident that the periodic arrangement of the dipole resonators, keeping fixed the total size of the tag, provides the highest cross-polar RCS. This because all the scatters provide a perfect polarization conversion and they are all in phase. On the contrary, the other two configurations do not guarantee this feature and each unit cell is characterized by a different reflection coefficient at a given frequency. The increase of the number of dipoles in the aperiodic arrangement provides a moderate increase of the RCS value but it strongly deteriorates the intelligibility of the signal by altering the maximum to minimum delta RCS value. In order to highlight the effect of the number of dipoles in the aperiodic configurations, the cross-polar RCS of the aperiodic configuration with three different number of dipoles is shown in Fig.~\ref{fig:RCS_vs_Ndipoles}.

\begin{figure}
    \centering
  \subfloat[][\textit{Periodic}]{%
       \includegraphics[width=0.45\linewidth]{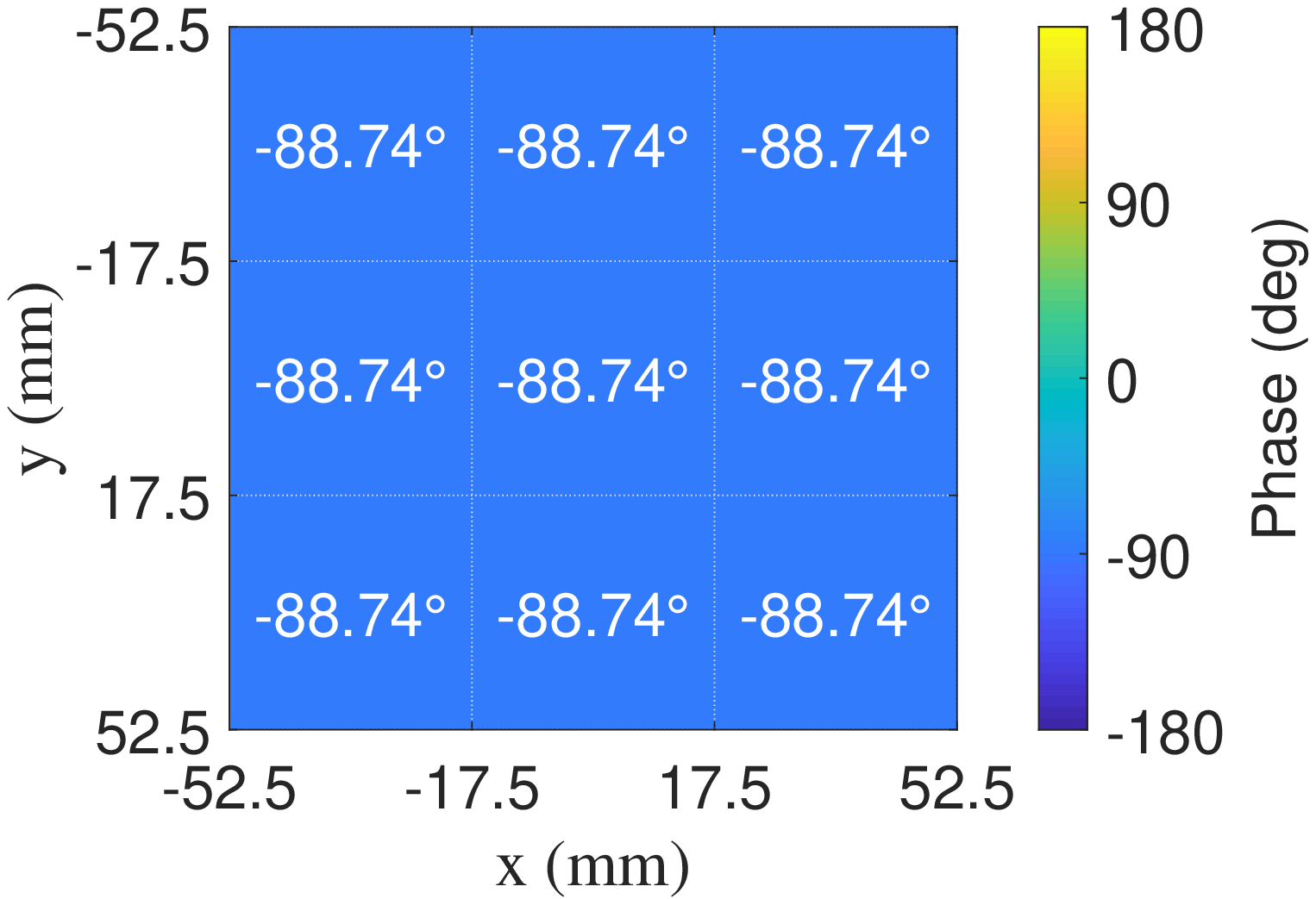}}
    \label{}\hfill
  \subfloat[][\textit{Periodic}]{%
        \includegraphics[width=0.45\linewidth]{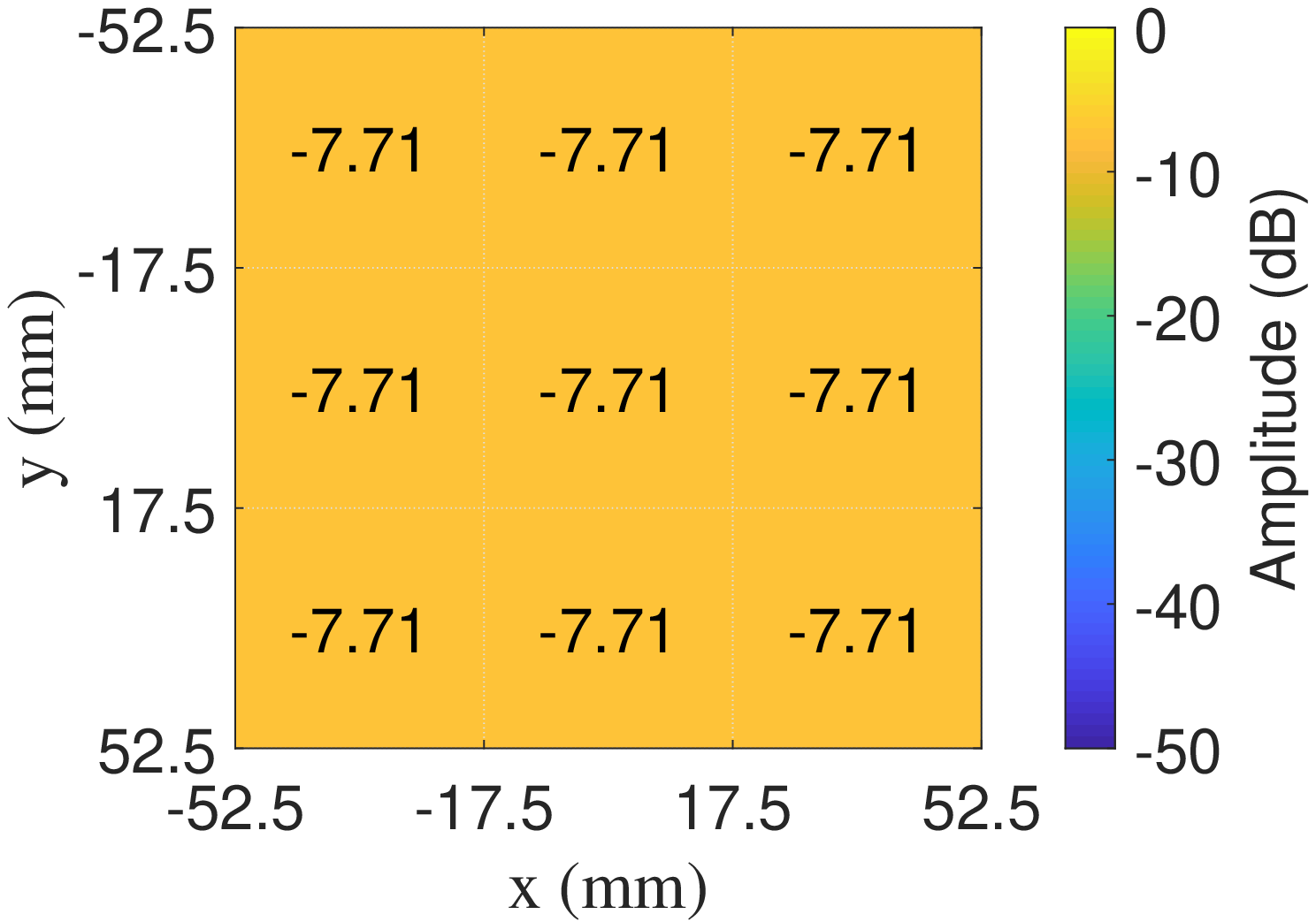}}
    \label{}\\
  \subfloat[][\textit{NonPeriodic-9}]{%
        \includegraphics[width=0.45\linewidth]{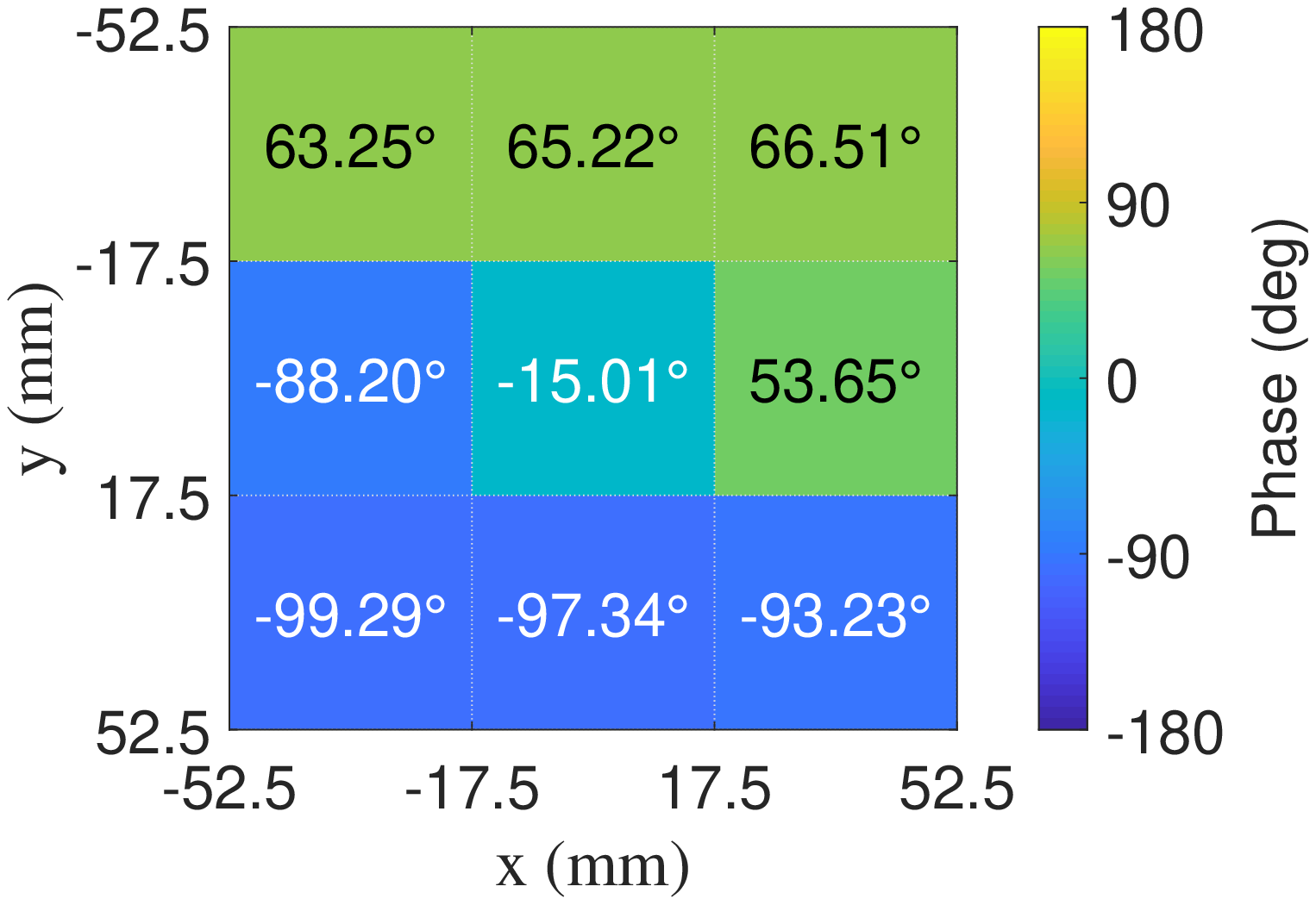}}
    \label{}\hfill
  \subfloat[][\textit{NonPeriodic-9}]{%
        \includegraphics[width=0.45\linewidth]{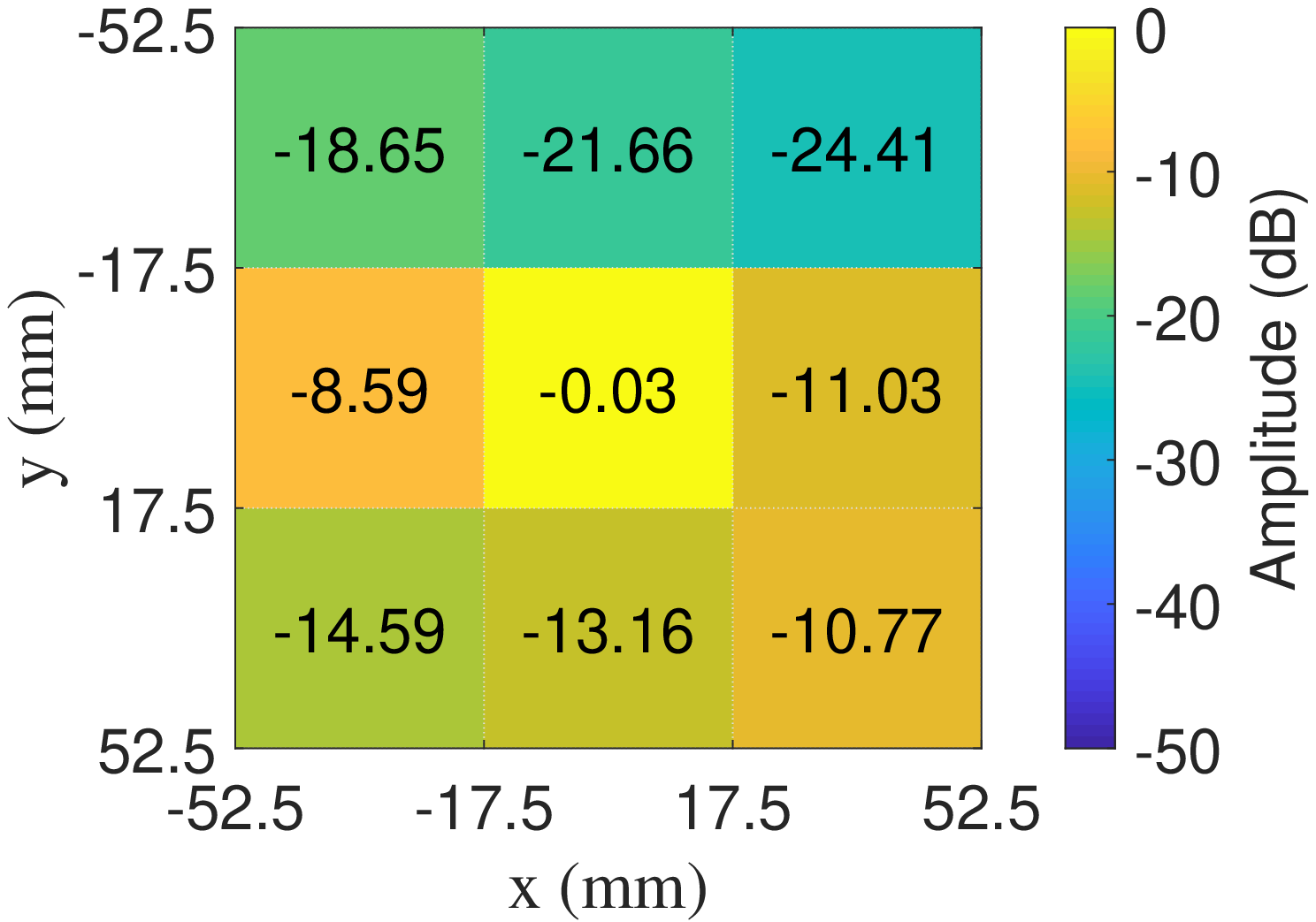}}
     \label{}\\
  \subfloat[][\textit{NonPeriodic-1}]{%
        \includegraphics[width=0.45\linewidth]{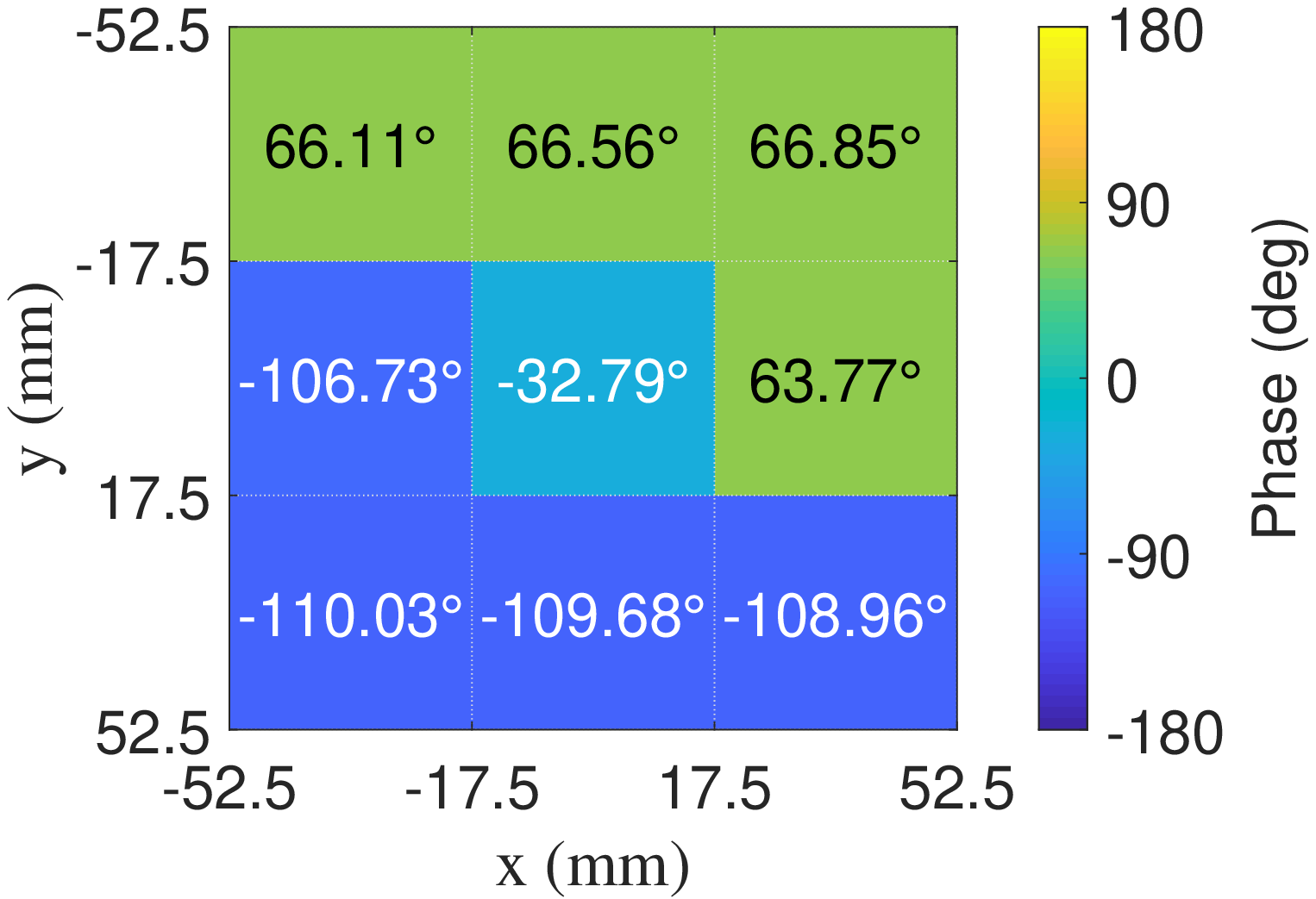}}
    \label{}\hfill
  \subfloat[][\textit{NonPeriodic-1}]{%
        \includegraphics[width=0.45\linewidth]{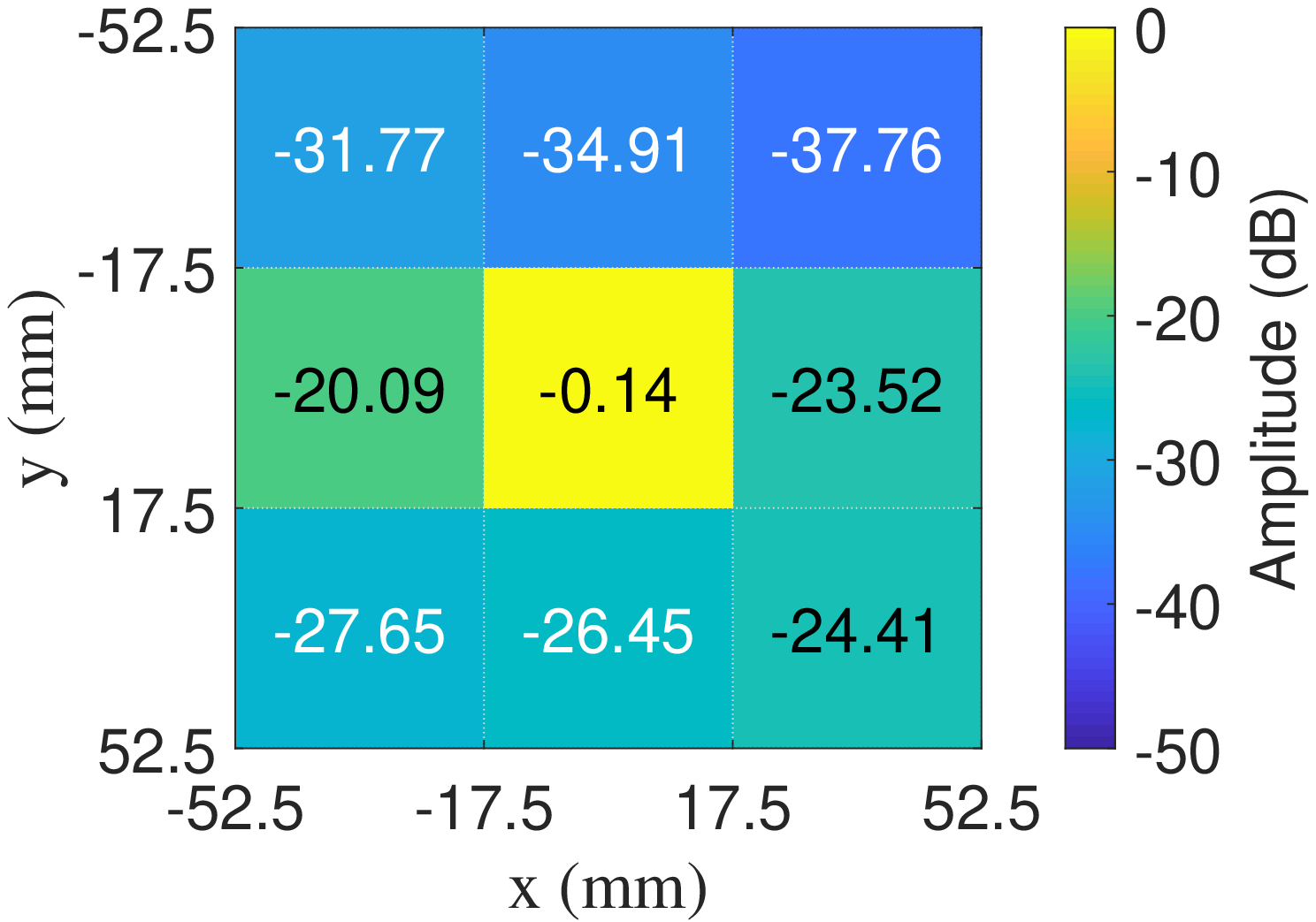}}
      \caption{Phase (a), (c), (e) and amplitude (b),(d),(f) distribution in the case of a $ 3\times3 $ dipole-based chipless tag arranged in a \textit{Periodic} and \textit{NonPeriodic-9} and \textit{NonPeriodic-1} configurations. The color plots are computed in correspondence of the fifth resonance peak of the tag: 4.7 GHz (\textit{Periodic}), 4.125 GHz (\textit{NonPeriodic-9}), 4.6 GHz (\textit{NonPeriodic-1}).}
\label{fig:amplitude_phase_distribution_dipoles} 
\end{figure}

\begin{figure} 
    \centering
  \subfloat[]{%
       \includegraphics[width=4.3cm,keepaspectratio]{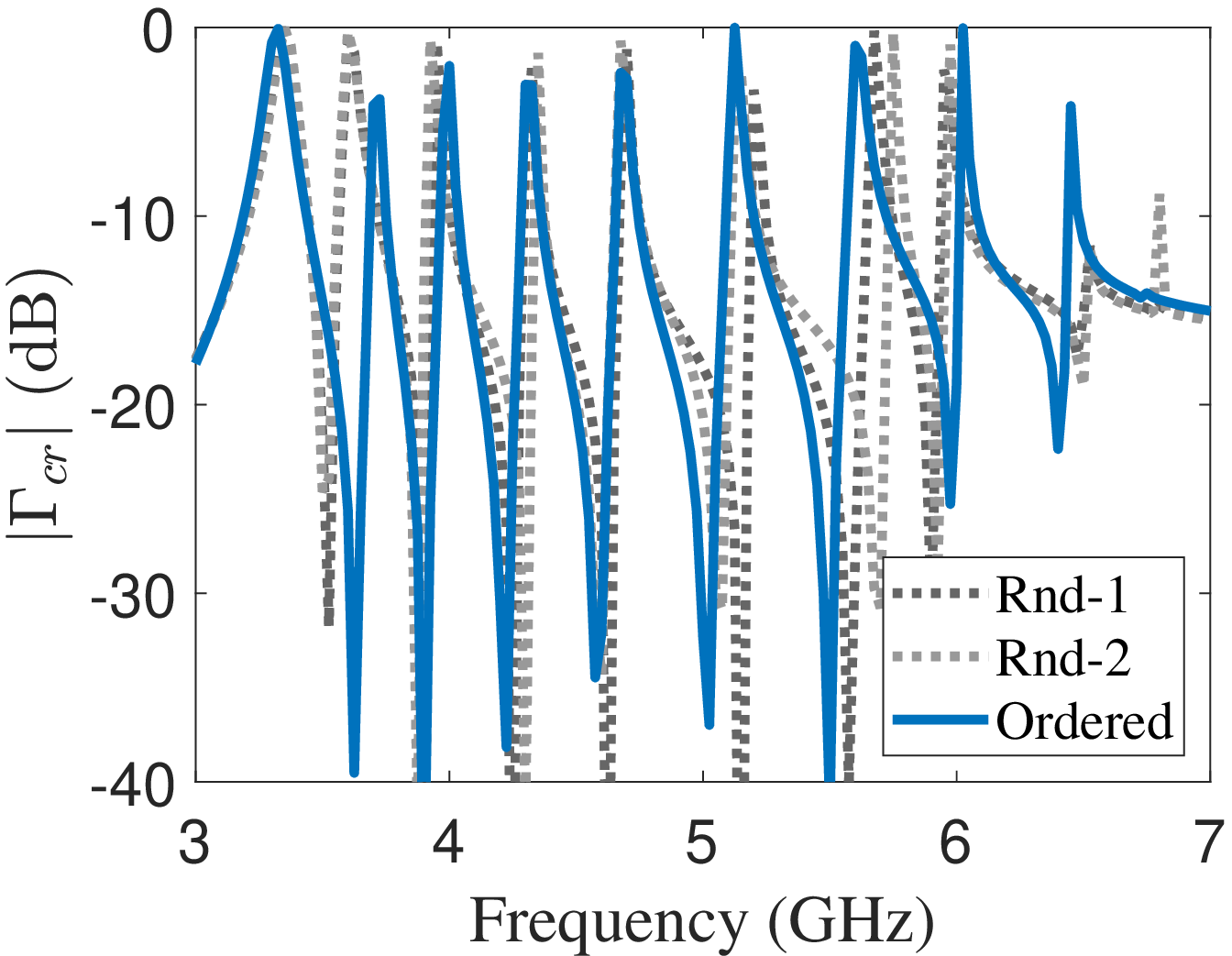}}
     \label{}\hfill
  \subfloat[]{%
        \includegraphics[width=4.3cm,keepaspectratio]{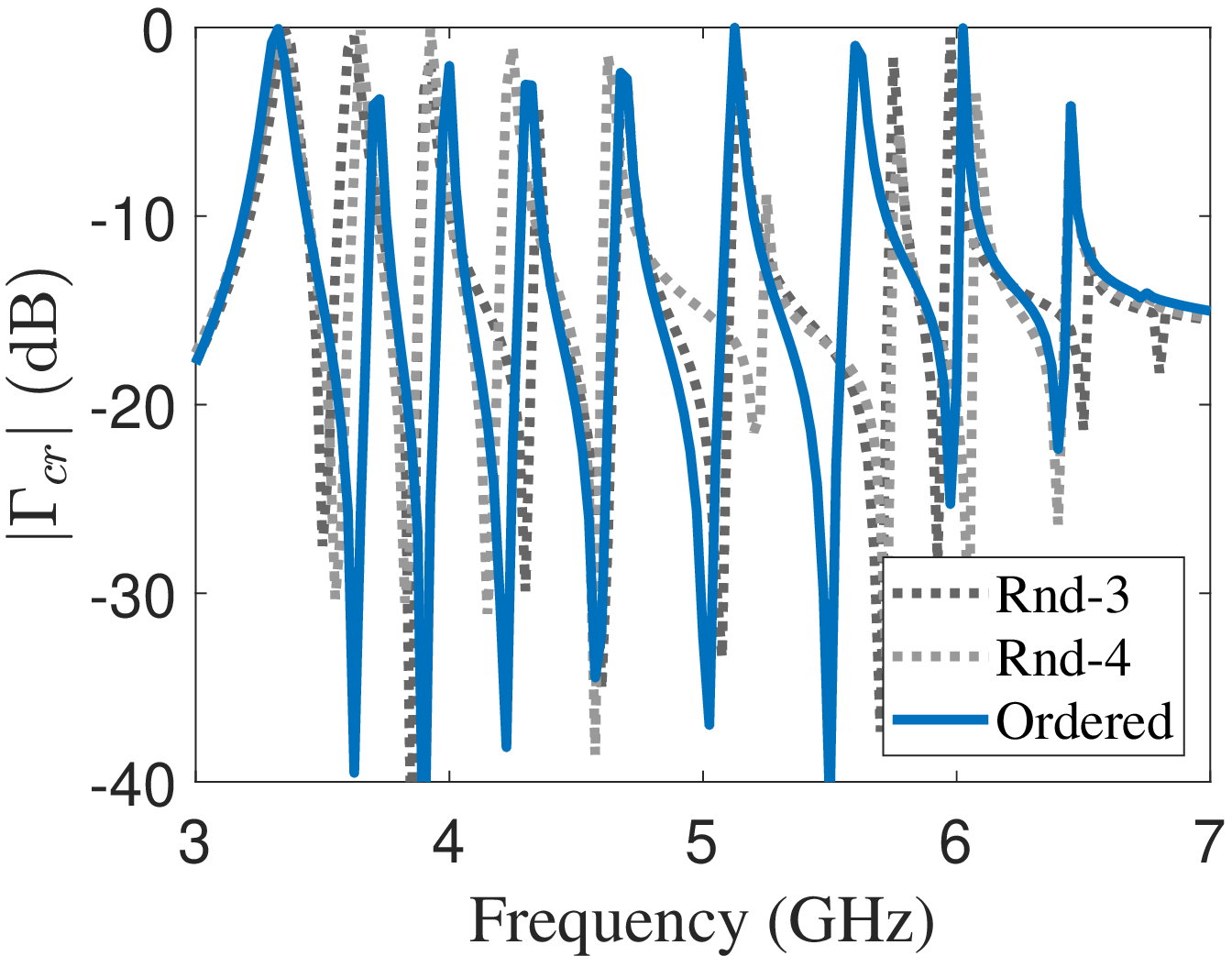}}
         \label{}\hfill
          \caption{Crosspolar RCS in the case of periodic configuration with different randomic disposition of dipoles.}
      \label{fig:reflection_dipoles_randomic} 
\end{figure}

In the non-periodic arrangement, the scatterers are characterized by a different amplitude and phase reflection coefficient and thus a destructive interference is obtained in the observed mono-static direction. This is evident by looking at the amplitude and phase profile of the reflection coefficient computed for the nine scatterers. The amplitude of the cross-polar reflection coefficients as a function of the frequency for the periodic and non-periodic configurations are reported in Fig.~\ref{fig:amplitude_reflection_coefficients}.

For the periodic configuration all the elements are characterized by the same cross-polar reflection, since the unit cell are simply repeated to occupy the total area of the tag whereas in the aperiodic configurations each element is characterized by a different reflection coefficient and thus only a portion of the bidimensional surface contributes to the RCS of the tag. Moreover there are also some differences between the \textit{NonPeriodic-9}, \textit{NonPeriodic-5} and \textit{NonPeriodic-1} configurations: the former is characterized by reflection coefficients with a lower Q-factor with respect to the latter configuration. This means that the scattering at a specific frequency in the \textit{NonPeriodic-9} configuration is strongly deteriorated by the neighboring scatterers which provides a relevant contribution which is not in phase with the main one. The same effect can be observed by plotting the reflection coefficient maps of the three analysed configurations for a specific frequency peak. In Fig.~\ref{fig:amplitude_phase_distribution_dipoles} the amplitude and phase reflection coefficients for the nine elements composing the tags are reported  at the frequency of the fifth resonant peak. It is evident that there is a destructive interference among the unit elements of the tag in the case of the aperiodic arrangement. 

\section{Disordered Dipoles arrangement}
In order to address the optimal disposition of dipoles in the tag, the ordered arrangement used in the previous configurations can be compared with the random arrangement. Indeed, the mutual coupling among the resonators can determine a different behaviour of the resonator which is difficult to control \textit{a-priori}. All the possible arrangements of the $N$ dipoles inside the tag or inside the unit cell of the periodic configuration can be computed as the factorial of the number of the dipoles `$N!$'. In the analysed case with 9 dipoles, the number of configurations is 362880. Consequently, a complete analysis would require a considerable amount of time. For this reason, we decided to analyse a subset of configurations and compare them with the reference configuration with the dipoles ordered from the longest to the shortest (\textit{Periodic} configuration). The random arrangements of the dipoles have been initially analysed for the \textit{Periodic} configuration. The cross-polar reflection coefficient of some of the analysed configurations are reported in Fig.~\ref{fig:reflection_dipoles_randomic}. The ideal configuration is the one with all the peaks characterized by a cross-polar reflection amplitude reaching 0~dB. As is evident from the sample configurations, the change of the dipole disposition determines a shift of the peaks in frequency and also a drop of some of them. Generally, the ordered configuration (named $Periodic$ in the previous sections) is a good compromise among the others. However, even this configuration may be theoretically improved in the peaks [2,3,4,5,9] (numbering them from low to high frequency) where the cross-polar reflection coefficient does not reach 0~dB. The analysed dispositions of the dipoles in the \textit{Periodic} and \textit{NonPeriodic} configurations are reported in Table \ref{tab:dipole_lenghts}.

\begin{table}[ht]
\caption{Length of the dipoles expressed in pixels. One pixel is equal to D/32 (1.09 mm), where D is the size of the elementary cell.}
\centering
\begin{tabular}{|l|c|c|c|c|c|c|c|c|c|}
\hline
 & \textbf{D1} & \textbf{D2} & \textbf{D3}& \textbf{D4} & \textbf{D5} & \textbf{D6}& \textbf{D7} & \textbf{D8} & \textbf{D9} \\
\hline
Rnd-1 & 18 & 26 & 20 & 28 & 16 & 17 & 24 & 30 & 22\\
\hline
Rnd-2 & 24 & 26 & 18 & 28 & 16 & 17 & 22 & 30 & 20\\
\hline
Rnd-3 & 30 & 24 & 18 & 26 & 16 & 17 & 22 & 28 & 20\\
\hline
Rnd-4 & 30 & 16 & 28 & 17 & 26 & 18 & 24 & 20 & 22\\
\hline
\end{tabular}
\label{tab:dipole_lenghts}
\end{table}

\begin{figure}
    \centering
    \subfloat[][\textit{Periodic}]{%
        \includegraphics[width=0.3\linewidth]{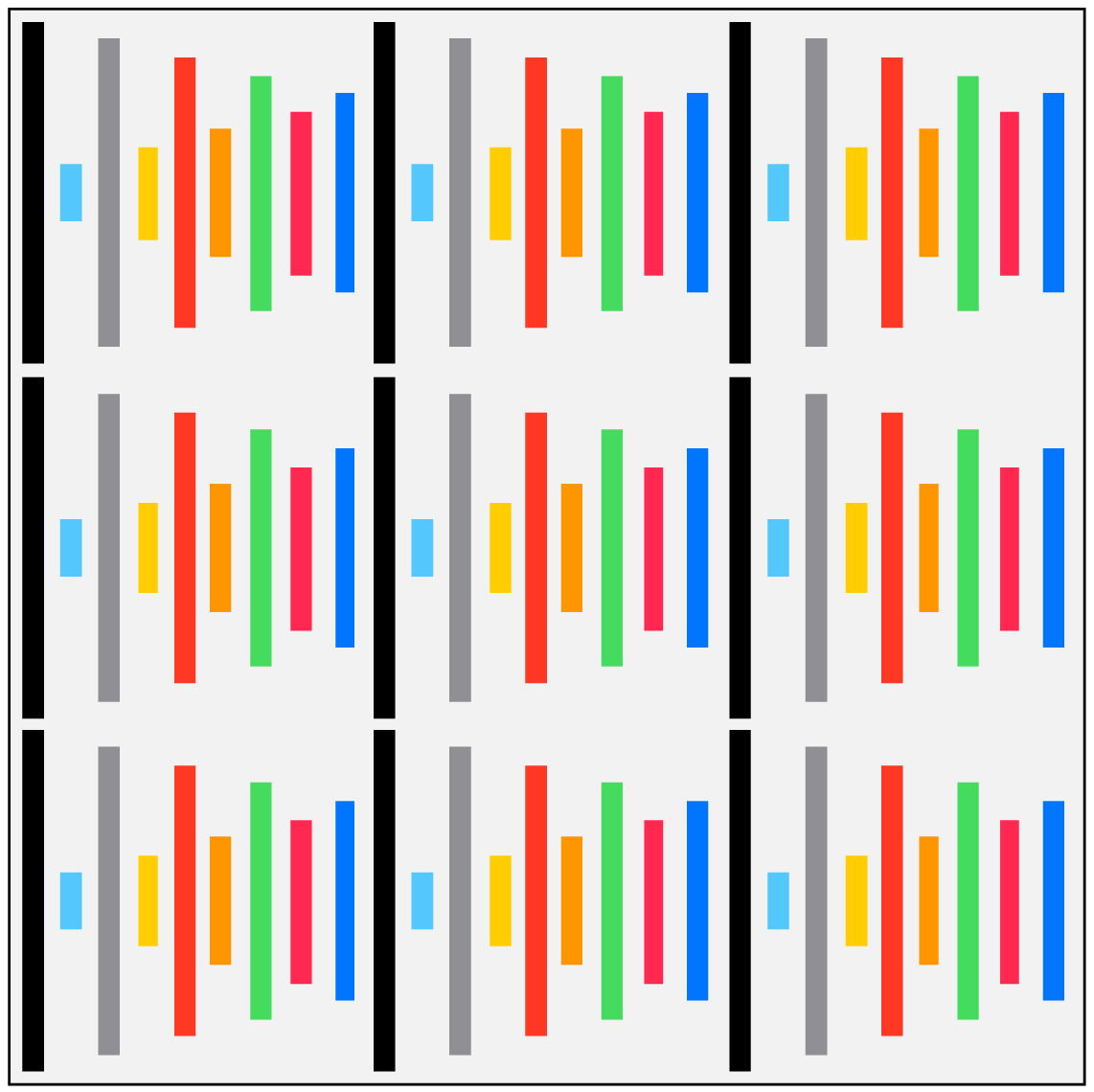}}
\label{}\hfill
  \subfloat[][\textit{NonPeriodic-9}]{%
        \includegraphics[width=0.3\linewidth]{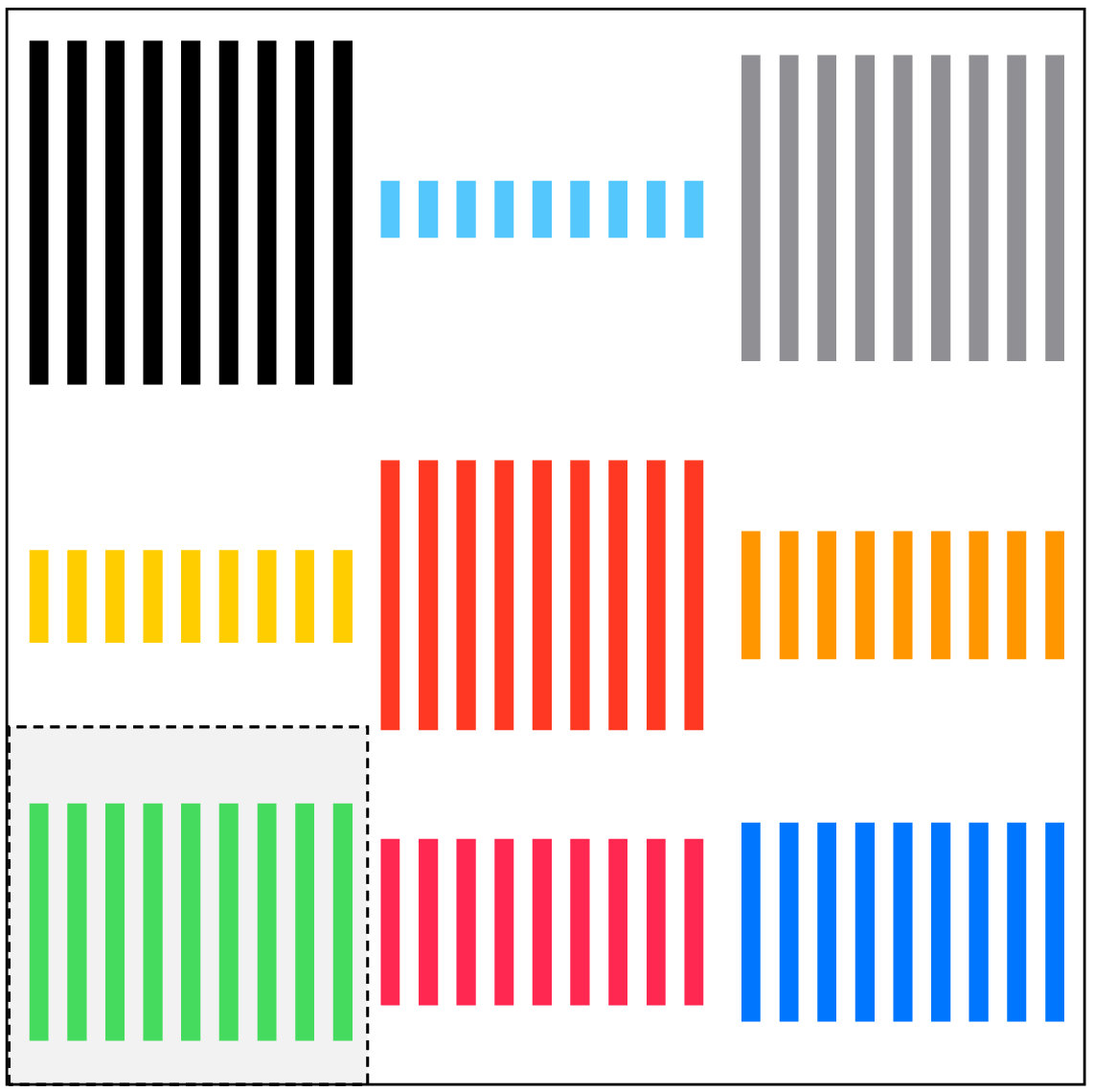}}
\label{}\hfill
  \subfloat[][\textit{NonPeriodic-1}]{%
        \includegraphics[width=0.3\linewidth]{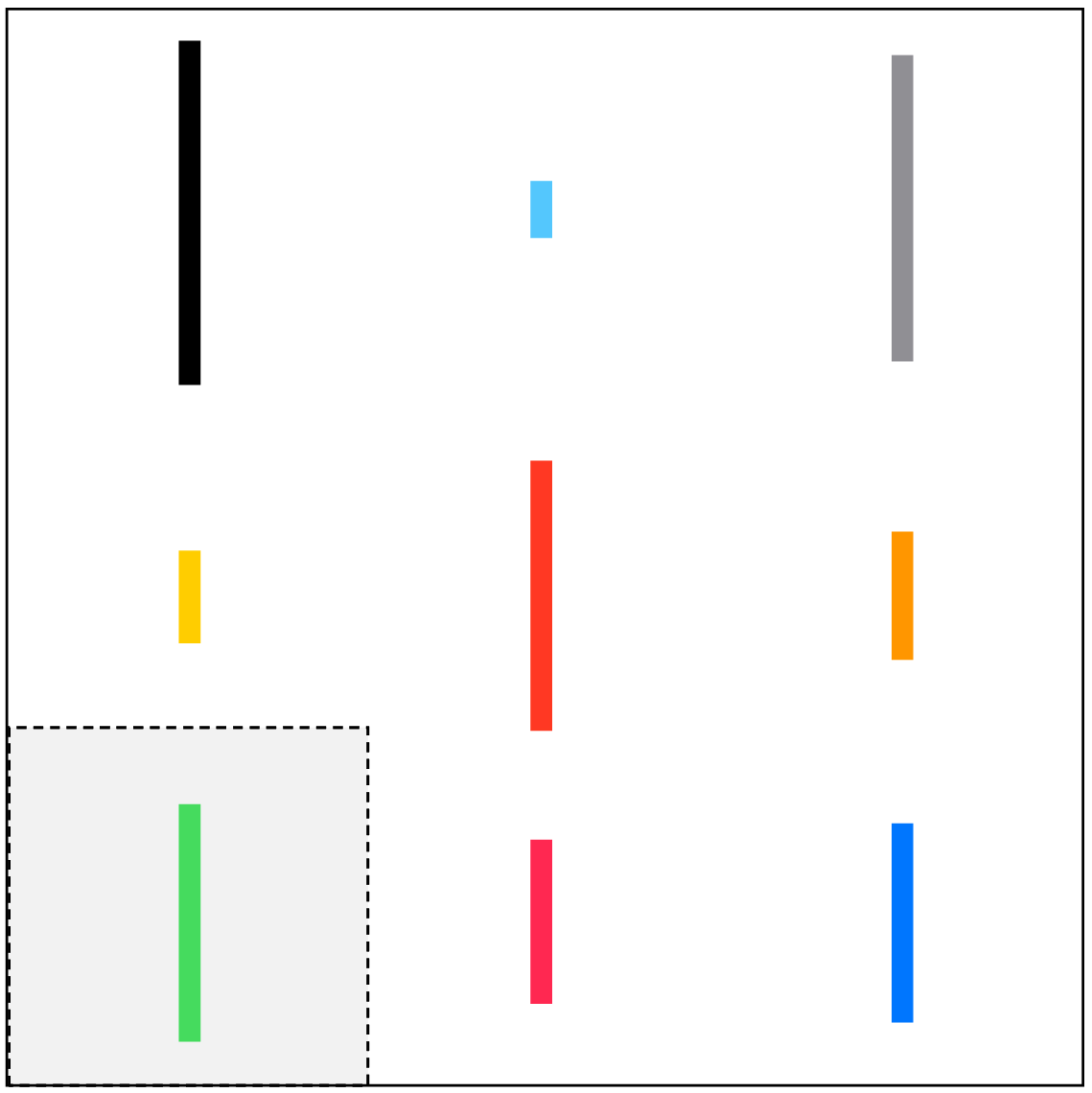}}
  \caption{Three different configurations of  dipole-based chipless tags characterized by the same physical area. (a) 81 $(9\times9)$ dipoles arranged in the \textit{Rnd-4} periodic configuration, (b) 81  and (c) 9 dipoles arranged in a \textit{Rnd-4} non-Periodic configuration. The area of the tag which provides an in-phase response is highlighted in grey.}
  \label{fig:conf_randmom} 
\end{figure}

\begin{figure}
    \centering
             \subfloat[][This method]{%
       \includegraphics[width=4.3cm,keepaspectratio]{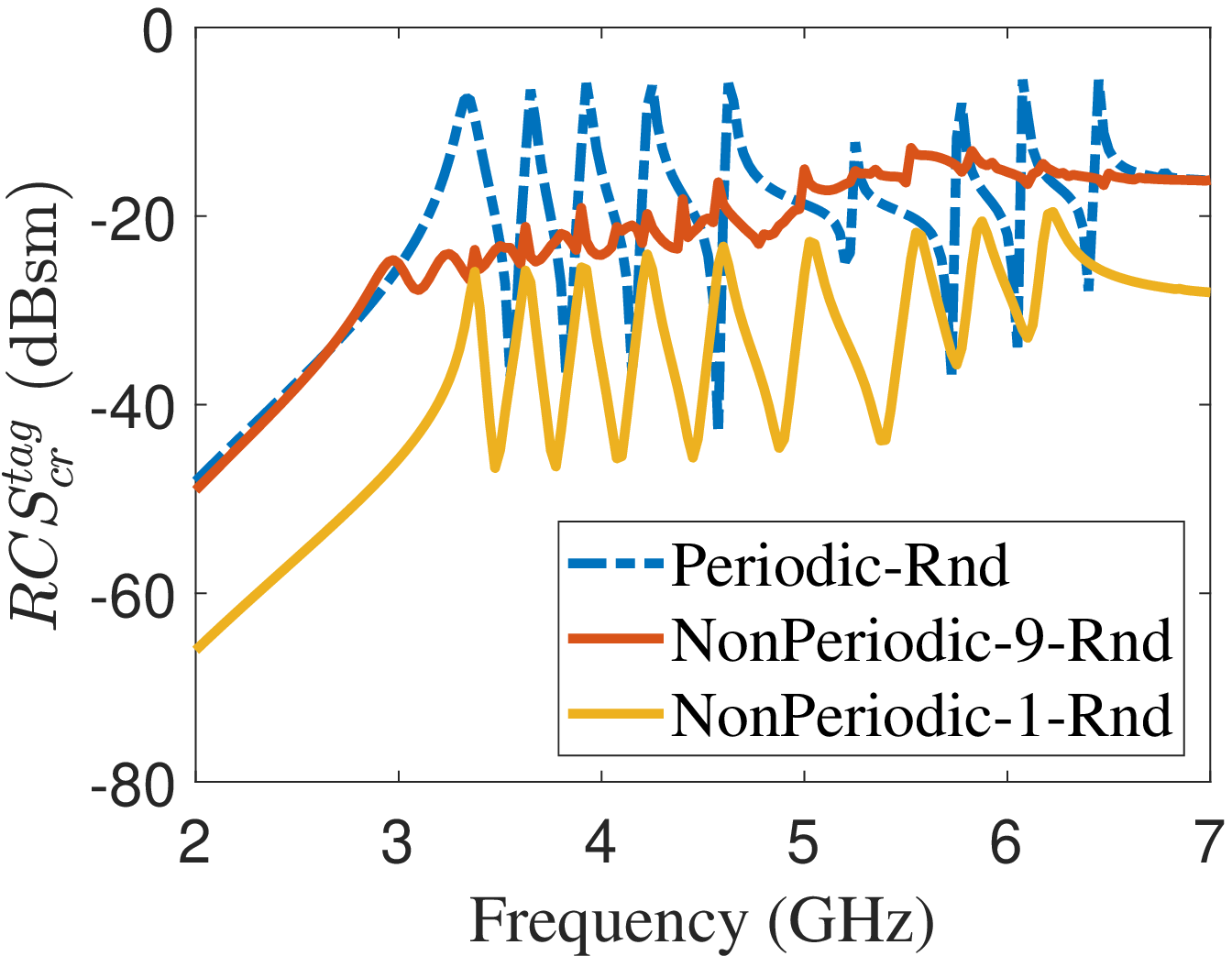}}
     \label{}\hfill
  \subfloat[][HFSS]{%
        \includegraphics[width=4.3cm,keepaspectratio]{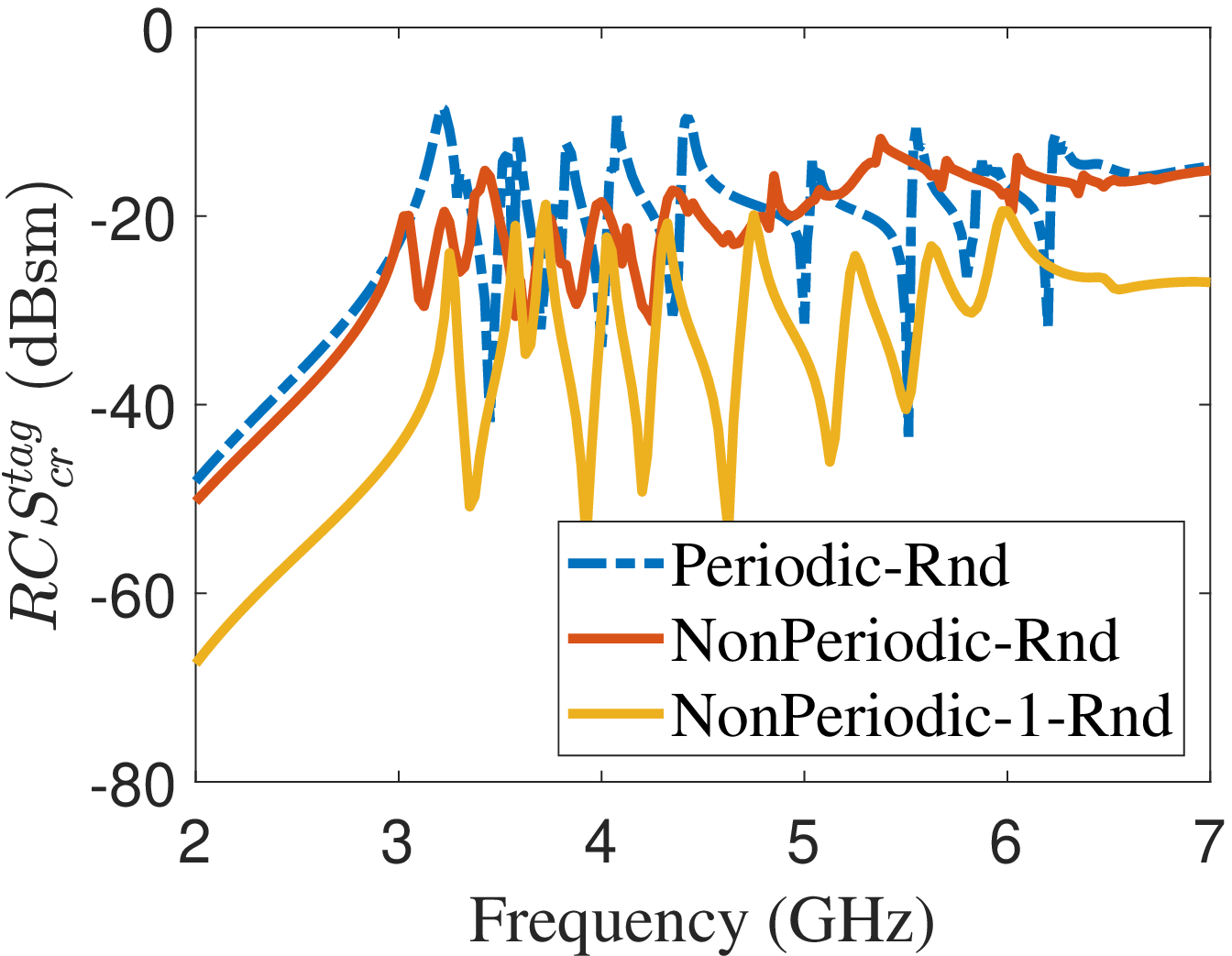}}
         \label{}\hfill
            \caption{Crosspolar RCS in the case of a dipole-based chipless tag randomly arranged in a periodic (\textit{Periodic-Rnd}) and non-Periodic configurations (\textit{NonPeriodic-9-Rnd} and \textit{NonPeriodic-1-Rnd}). (a) Proposed approach, (b) Ansys HFSS. The substrate is 2 mm thick and it is characterized by a dielectric permettivity equal to 2.08.}
      \label{fig:RCS_dipoles_random} 
\end{figure}

One of the randomic configurations previously analyzed, i.e. \textit{Rnd-4}, has been further analyzed also for the \textit{NonPeriodic-9} and \textit{NonPeriodic-1} configurations.   
The disposition of the dipoles according to the arrangement named \textit{Rnd-4} in Table \ref{tab:dipole_lenghts} are shown in Fig.~\ref{fig:conf_randmom} both for the periodic and aperiodic configurations. The cross-polar RCS of the three investigated configurations are reported in Fig.~\ref{fig:RCS_dipoles_random}. The RCS has been computed by the proposed semi-analytical formulation and by using a full-wave solver (Ansys HFSS). The general behavior of the different types of arrangements are in agreement and thus the validity of the proposed RCS computation approach is confirmed. As in the previous ordered disposition, the periodic arrangement of the dipole resonators, keeping the total size of the tag fixed, provides the highest cross-polar RCS. However, the random disposition determines a decrease of the amplitude of the sixth and seventh  peaks because of the mutual coupling between the resonators.

\section{Measurements}  
\begin{figure}
    \centering
  \subfloat[][\textit{Periodic}]{%
       \includegraphics[width=3.5cm,keepaspectratio]{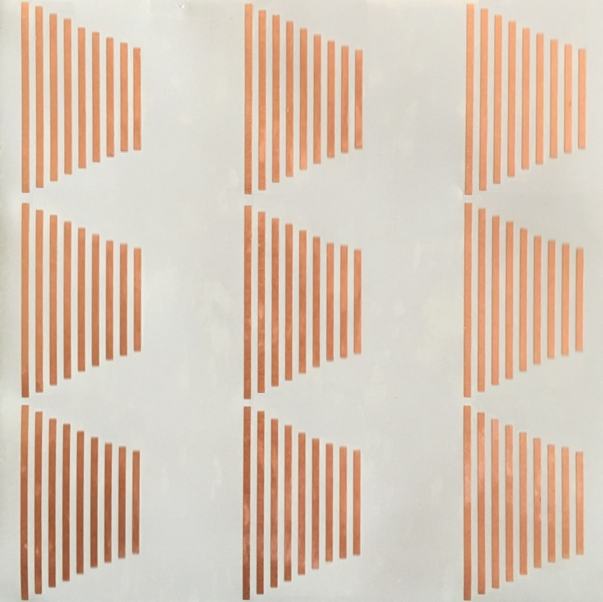}}
     \label{}\hspace{15pt}
  \subfloat[][\textit{NonPeriodic-9}]{%
        \includegraphics[width=3.5cm,keepaspectratio]{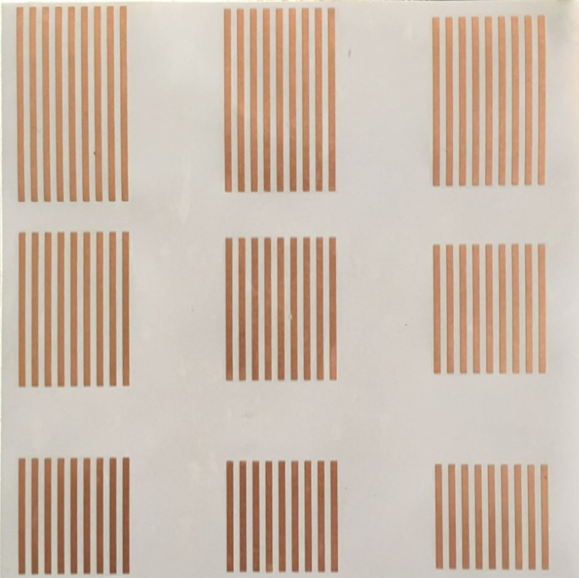}}
         \label{}
      \caption{Pictures of the fabricated tag prototypes: (a) \textit{Periodic}, (b) \textit{NonPeriodic-9}}
      \label{fig:Prototipi} 
\end{figure}

\begin{figure}
    \centering
       \includegraphics[width=8.8cm,height=4.4cm,keepaspectratio]{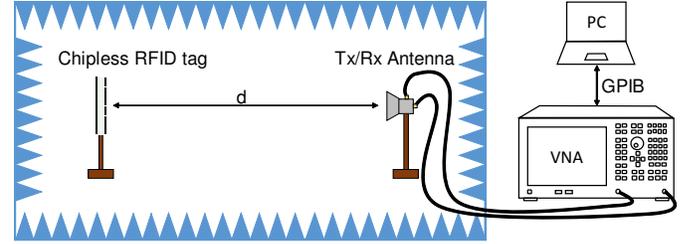} 
       \caption{Pictorial representation of the measurement setup.}
      \label{fig:setup}    
\end{figure}

\begin{figure}
    \centering
        \includegraphics[width=7cm,keepaspectratio]{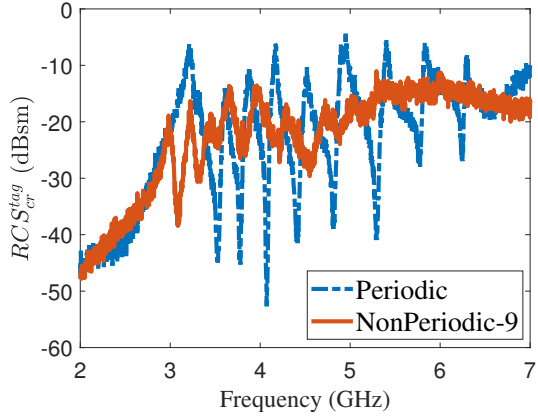}   
      \caption{Measured RCS of the fabricated prototypes shown in Fig.~\ref{fig:Prototipi}.}
      \label{fig:Misure} 
\end{figure}

In order to verify the accuracy of the predictions obtained with both by the proposed  approach for computing the RCS and by the full-wave simulations, the prototypes of two dipole-based depolarizing chipless tags have been fabricated. The picture of the fabricated prototypes and the measured cross-polar RCS of the two tags is reported in Fig.~\ref{fig:Prototipi}. The RCS of the fabricated prototypes has been measured in an anechoic environment with a single duel polarized wideband horn antenna. The RCS level has been computed by using as a reference a metallic plate of the same dimension of the tags. A pictorial representation of the measurement setup is shown in Fig.~\ref{fig:setup}. The cross-polar measured RCS level $RCS^{tag}_{cr}$ is shown Fig.~\ref{fig:Misure}. As predicted by numerical simulations, the cross-polar RCS of the periodic tag is characterized by a much higher RCS value with respect to the non-periodic configuration. Moreover, the periodic configuration guarantees a much higher intelligibility of the RCS peaks since, at a resonance frequency, all the zones of the tag respond in phase as shown by the grey area in Fig.~\ref{fig:in_phase_area}(a). On the contrary, by using the non-periodic arrangement of the resonators, only a small part of the tag responds at a specific frequency (grey area in Fig.~\ref{fig:in_phase_area}(b),(c)), whereas the other zones of the tag create a destructive interference that deteriorates the quality of the signal. The non-periodic configuration has however a couple implicit advantages with respect to the periodic one. The first is that the resonators are less coupled and thus the removal of a specific frequency peak introduces a limited frequency shift in the remaining peaks. The second is that it is possible to identify which part of the tag is responding to an RF interrogation by associating a specific zone of it to a specific frequency. This feature can be usefully for realizing a positioning sensing tag \cite{barbot2017gesture}. 

\section{BER performance}  
\begin{figure}
\centering
\centerline{\includegraphics[width=9cm,keepaspectratio]
{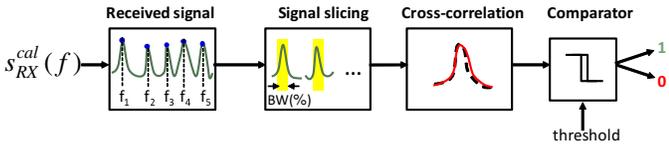}}
\caption{Schematic representation of the processing at the receiver and the decision method.}
\label{fig:Rx_scheme}
\end{figure}

As previously stated, the signal at the receiver is composed by the desired signal and by additional undesired contributions. In order to detect if the bit sequence is correct or not, the noisy signal at the receiver is subdivided (sliced) into $N$ signals around the resonance frequencies used to encode information (a percentage bandwidth $BW$ around each peak is retained) and each of them is compared with the ideal version of the signal. A correlation function between the perturbed signal and the ideal one is carried out for each bit. The presence or the absence of a bit in a certain frequency band is evaluated by computing the correlation coefficient \textit{R} between the unperturbed  signal and the noisy one. A correlation coefficient \textit{R} higher than $0.8$ codifies the bit ``1" (presence of the bit). Conversely, a condition \textit{R} lower than $0.8$ codifies the bit ``0" (absence of the bit). Typically, a strong cross correlation between two variables results in a correlation coefficient larger than 0.7 \cite{moore2007basic}. We selected a threshold of 0.8 for deciding if the bit is ``0" or ``1" thus setting a threshold slightly higher than the one which commonly identifies a strong correlation. Based on this decision method, the error probability is then computed for different levels of transmitted power over the clutter contribution according to  eq.~(\ref{eq:rec_signal_approx}). The block diagram of the receiver is depicted in Fig.~\ref{fig:Rx_scheme}. The Bit Error Rate (BER) is evaluated by using a set of Montecarlo simulations with $ N=10^6 $ realizations of the process.

\begin{figure}
\centering
\centerline{\includegraphics[width=9cm,keepaspectratio]
{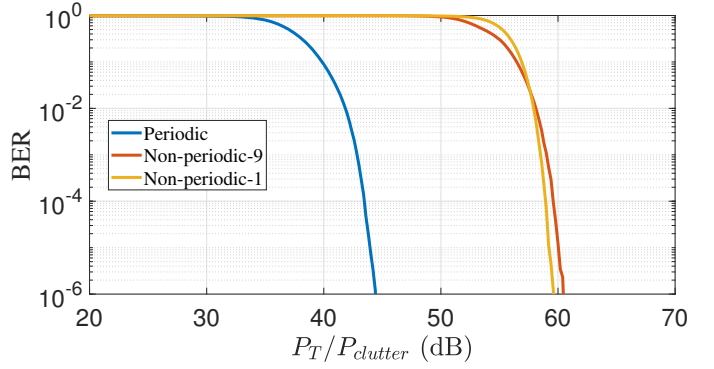}}
\caption{BER as a function of the transmitted power over the clutter power ($ P_T/P_{clutter} $) in the case of a $ 3\times3 $ dipoles-based chipless tag arranged in a periodic and non-periodic configuration. $P_T=1$ W, $d= 50$ cm.}
\label{fig:BER}
\end{figure}

\begin{figure}[h]
    \centering
  \subfloat[]{%
       \includegraphics[width=4.3cm,keepaspectratio]{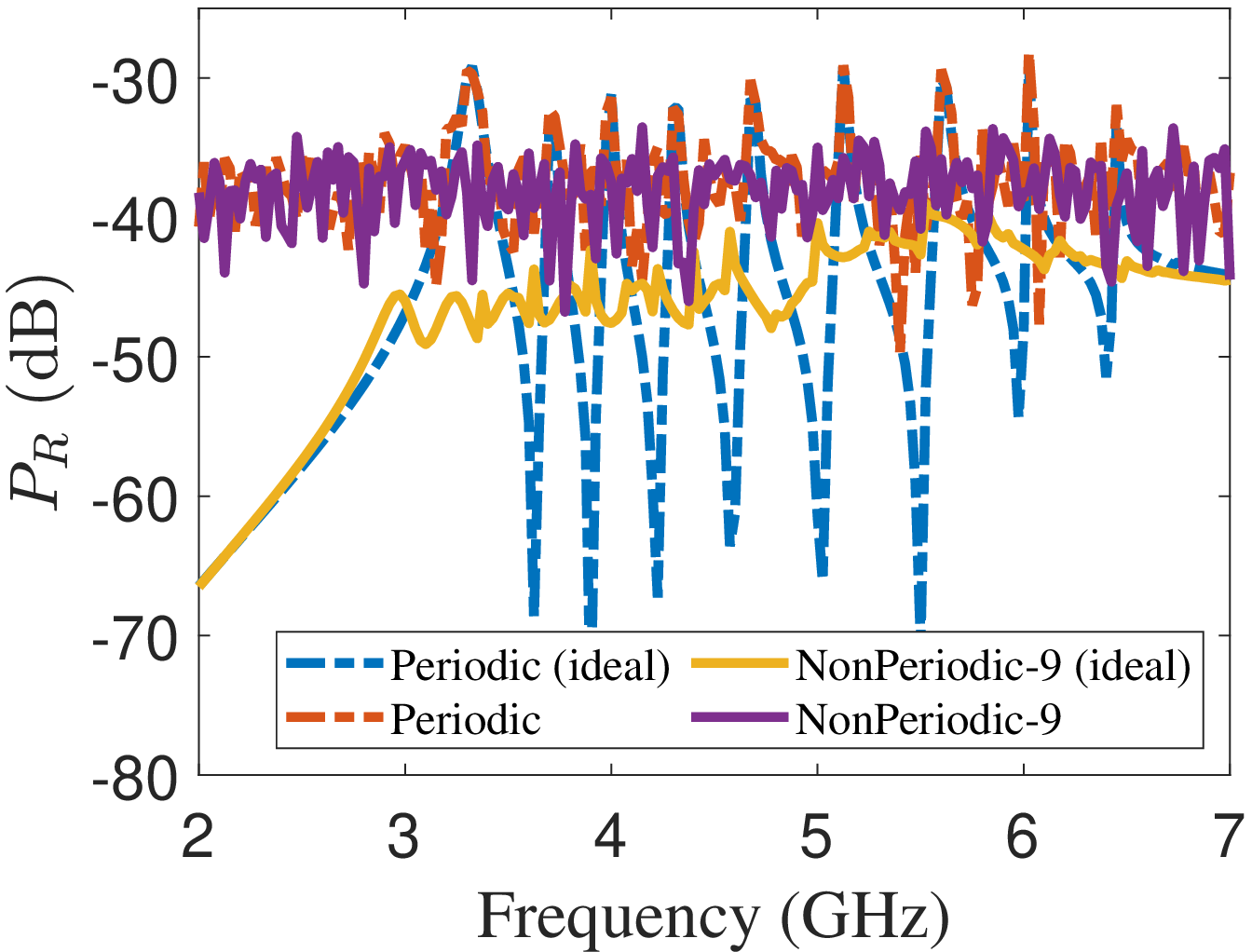}}
     \label{}\hfill
  \subfloat[]{%
        \includegraphics[width=4.3cm,keepaspectratio]{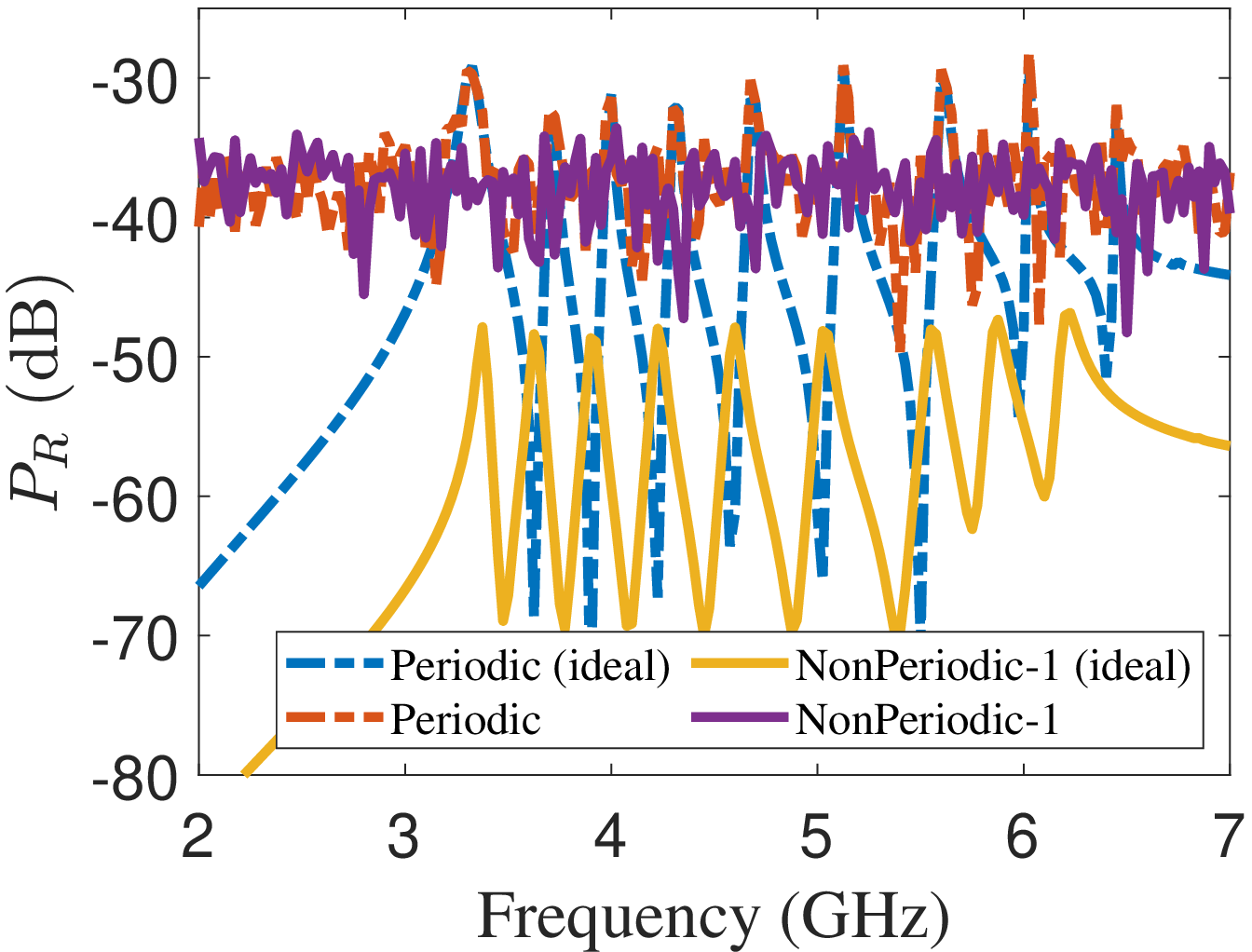}}
         \label{}\hfill 
      \caption{Received power ($P_R$) computed with radar equation (eq.~(\ref{eq:PR})) with and without (ideal) clutter: comparison between (a) \textit{Periodic} and \textit{NonPeriodic-9} tag; (b) \textit{Periodic} and \textit{NonPeriodic-1} tag.}
      \label{fig:signal_received} 
\end{figure}

The BER as a function of the transmitted power  ($ P_T=1 $~W) over the clutter contribution ($ P_{clutter} $) in the three cases analysed in section \ref{sec:RCS} (\textit{Periodic}, \textit{NonPeriodic-9}, \textit{NonPeriodic-1}). The estimated BER for these examples is reported in Fig.~\ref{fig:BER}. It is apparent that the probability of error mostly depends on the RCS value of the tag and it is therefore crucial to maximize this parameter in order to achieve a good detection of the tag. The waveform of the signal (a sharp and high frequency peak) is also important but it plays a role only if the received power exceeds the level of the residual noise. The effect of the signal waveform is evident by comparing at the \textit{NonPeriodic-9} and \textit{NonPeriodic-1} cases. The \textit{NonPeriodic-9} configuration shows better performance with $P_T/P_{clutter} < 57.6$~dB. However, the \textit{NonPeriodic-1} starts having better performance with $P_T/P_{clutter}>57.6$~dB because, despite the lower RCS with respect to the \textit{NonPeriodic-9} configuration, the signal waveform is more intelligible.
In communication systems, the probability of error can be expressed in a closed form since the BER is evaluated starting from the received signal level over the AWGN noise level \cite{fuschini2008}. However, in our case, we have expressed the BER as a function of the transmitted power over the clutter level for highlighting the effect of the RCS. The dependence of the BER on the RCS level is highlighted by the radar equation in (\ref{eq:PR}) which relates the received power level to  the RCS level. In order to better clarify this dependence, we have presented a couple of examples of received signal level in presence of the periodic tag and the aperiodic tag in presence of clutter. As shown in Fig.~\ref{fig:signal_received}, the received signal with the periodic tag is above the clutter floor level whereas the signal received with the aperiodic tag configurations is below the clutter level and therefore the bit sequence is not readable. 

\section{Conclusion}
The impact of the resonators arrangement on the system level performance of a chipless RFID communication system has been investigated. The cross-polar RCS of the tag is considerably affected not only by the shape of the resonant elements but also by their arrangement. The RCS of chipless tags comprising dipole resonators has been analytically calculated relying on planar reflectarray theory. Each resonant element has been modelled as a point scatterer characterized by a complex reflection coefficient. Considering a tag with a certain fixed physical area, it has been shown that a periodic arrangement of the resonators provides an higher level of RCS with respect to a non-periodic one. Indeed, according to array theory, a periodic arrangement of the resonators can be modelled as a planar array with elements radiating with a uniform amplitude tapering and the same phase at each working frequency of the tag. Conversely, a non-periodic arrangement results in a planar array of radiating elements with both a non-uniform amplitude and phase distribution. Experimental measurements of a chipless tag based on resonant dipoles arranged in a periodic and non-periodic configuration exhibit a good agreement with theoretical results. Finally, the strict dependency of system performance of the chipless RFID communication system, evaluated in terms of BER in a standardized scenario, has been demonstrated. The BER of the system is inversely proportional to the RCS of the tag. The BER is also influenced by the shape of the scattered signal in terms of peak-to-peak RCS. It has been shown that the BER represents the most meaningful approach for comparing the performance of different tags, instead of commonly used figures of merit such as \SI[detect-weight]{}{bit/cm^2} or \SI[detect-weight]{}{bit/Hz}. Indeed, the increase of encoded information in a certain physical area or in a certain bandwidth is useful only if the information can be correctly decoded.

\appendices
\section{Polarization Conversion Mechanism}
  \label{Appendix-A}
  
\begin{figure} 
    \centering
  \subfloat[]{%
       \includegraphics[width=4.2cm,height=4.2cm,keepaspectratio]{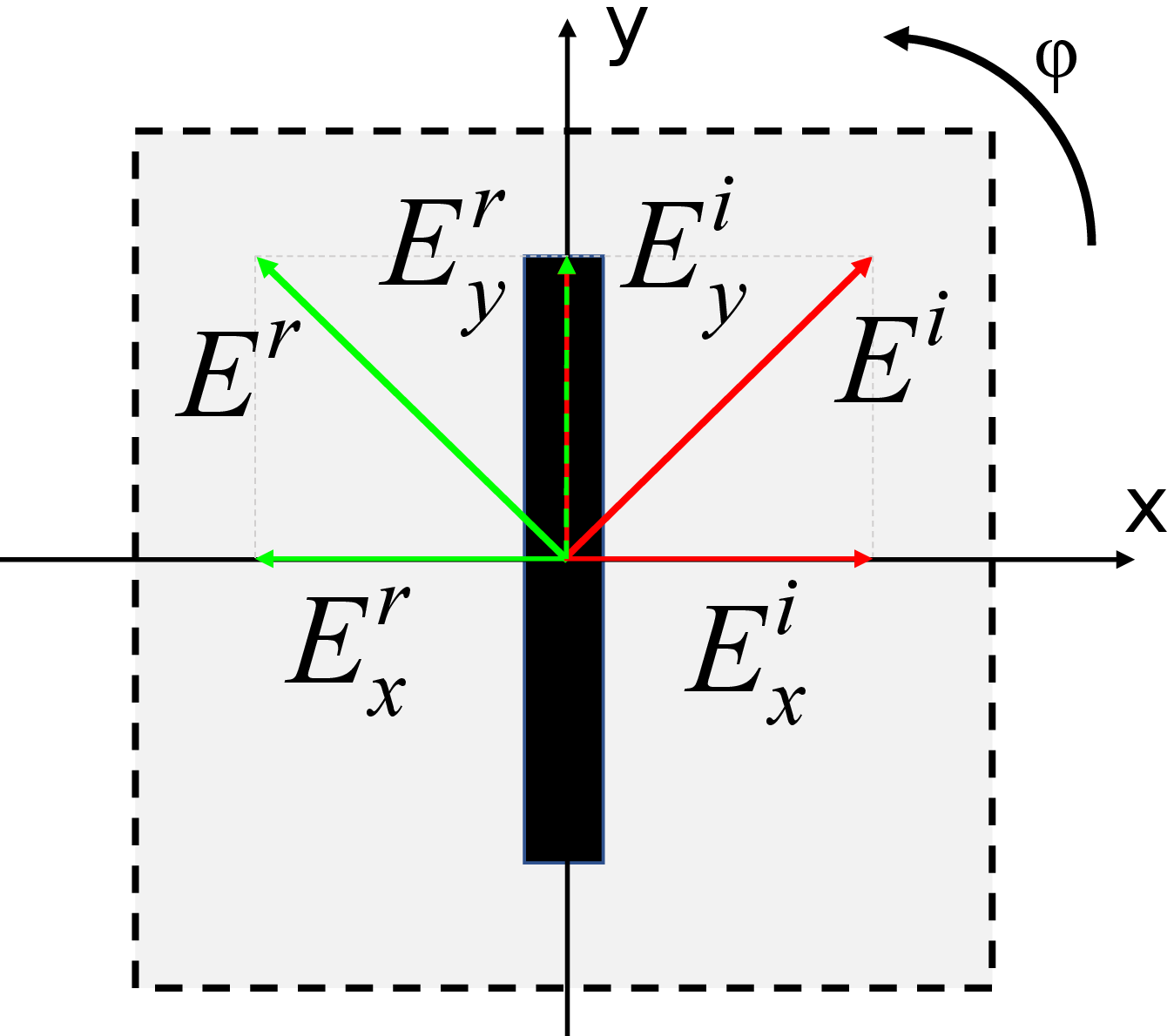}}
    \hfill
  \subfloat[]{%
        \includegraphics[width=4.4cm,height=4.4cm,keepaspectratio]{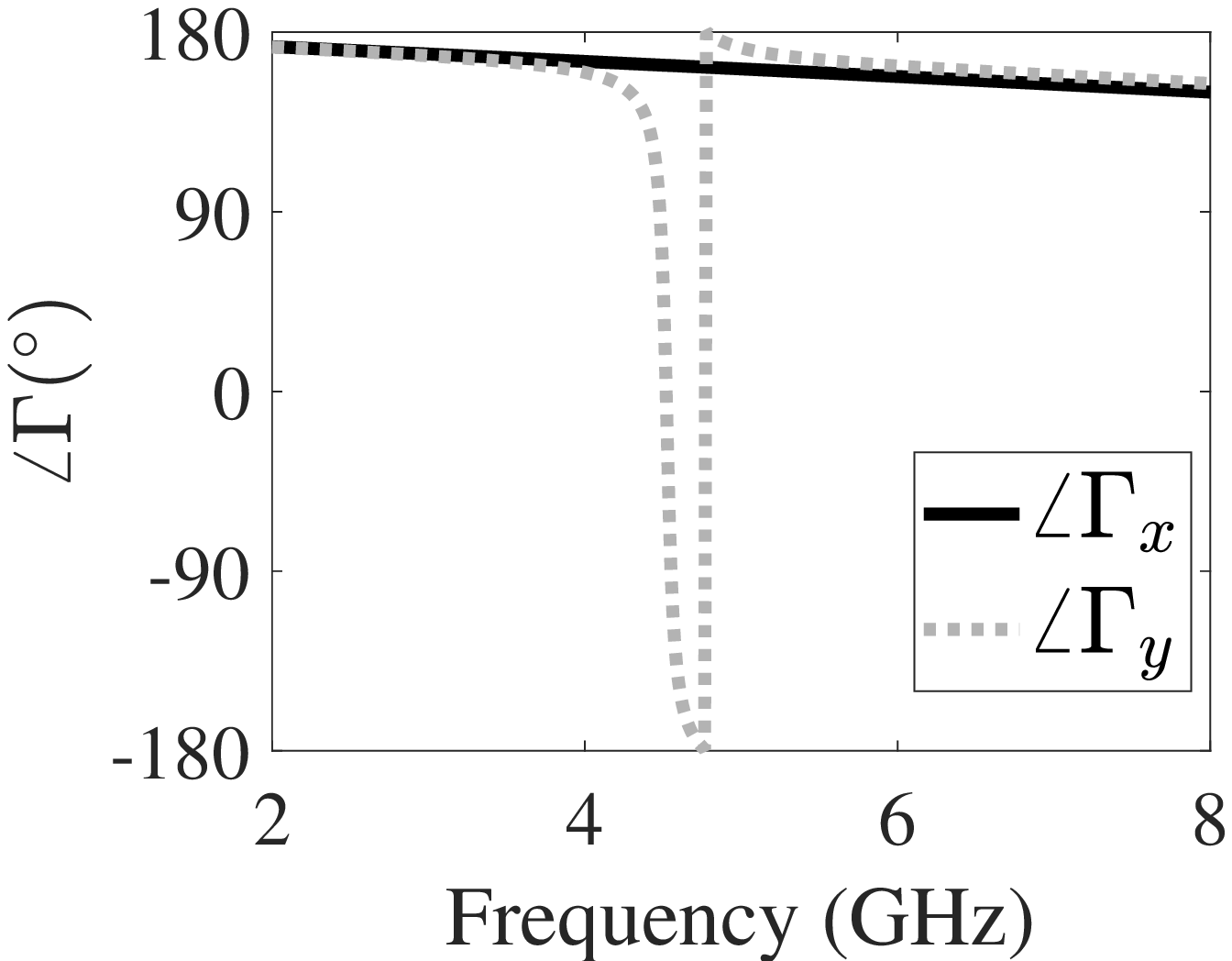}}
    \label{fig:FSSeqCirc} 
     \hfill
  \subfloat[]{%
        \includegraphics[width=4.4cm,height=4.4cm,keepaspectratio]{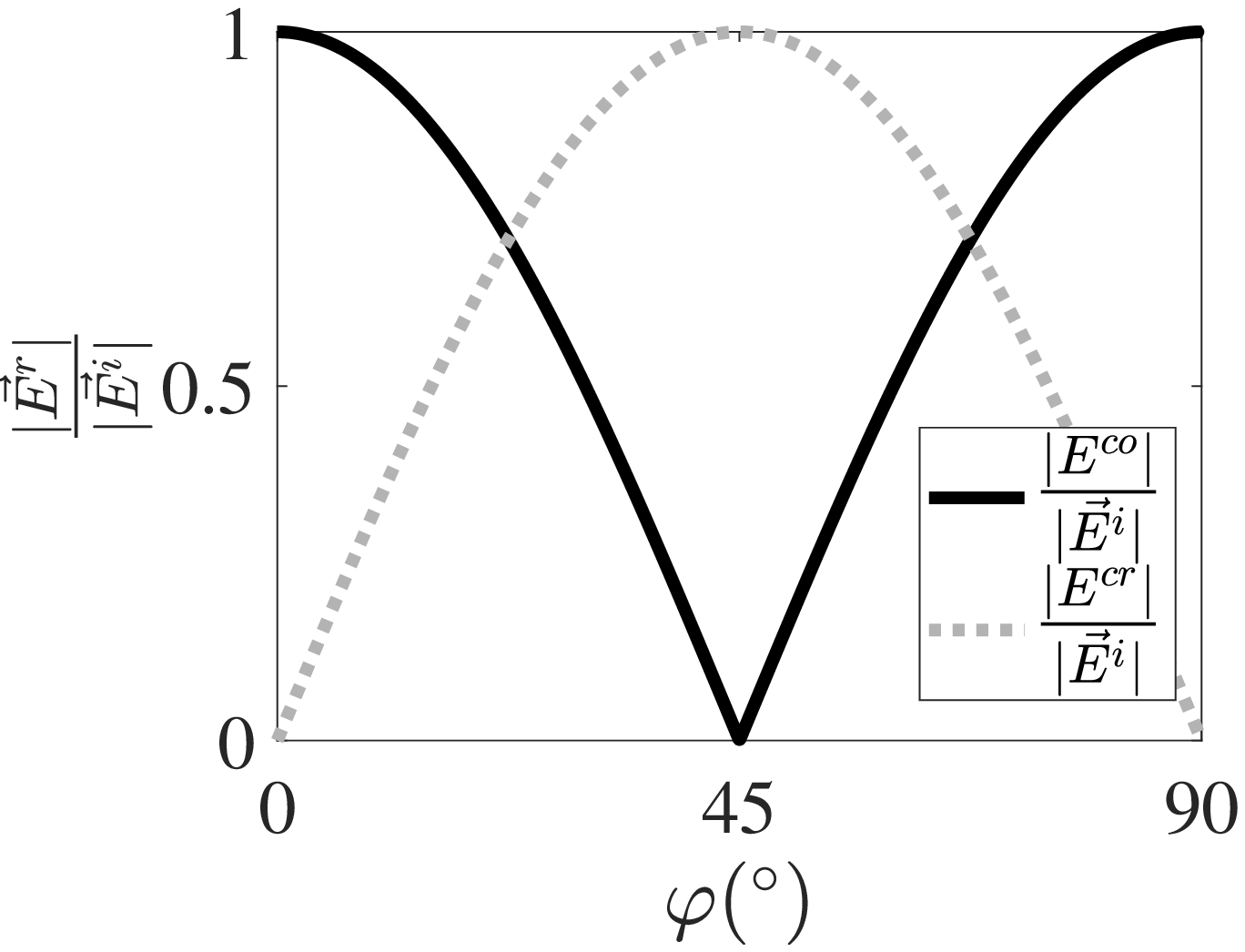}}
  \caption{(a) Incident and reflected electric field decomposed in the $ (x,y) $-coordinate system; (b) Phase of the reflection coefficient as a function of frequency; (c) Co-polar and cross-polar components of the reflected electric field normalized to the magnitude of the incident electric field as a function of the rotation angle $ \varphi $.}
    \label{fig:PolConvDipole}
\end{figure}

The working principle of a polarization converting surface based on dipole resonators printed on a ground plane is analysed. A dipole resonator placed at a certain distance from a metallic ground plane behaves as a perfect polarization converted at a single frequency when an electric field impinges at $ 45^\circ $ with respect to the orientation of the dipole as depicted in Fig.~\ref{fig:PolConvDipole}(a). The working principle of the polarization converter can be explained by decomposing the $ 45^\circ $ incident field ($ E^i $) in two identical vectors $ E_x^i $ and $ E_y^i $ along $ x $ and $ y $ axes as depicted in Fig.~\ref{fig:PolConvDipole}(b):

\begin{equation}
{\vec E^{i}} = {E_0}\hat \varphi={E_0}\sin \left( {\frac{\pi}{4}} \right)\hat x + {E_0}\cos \left( \frac{\pi}{4} \right)\hat y.
\end{equation}

The reflected electric field  $\vec E^{r}$ can be obtained from the incident electric field  $\vec E^{i}$ as follows:

\begin{equation}
{\vec E^{r}} = {\Gamma_x}\,{E_0}\sin \left( {{\varphi}} \right)\hat x + {\Gamma_y}\, {E_0}\cos \left( {{\varphi}} \right)\hat y.
\end{equation}

As shown in Fig.~\ref{fig:PolConvDipole}(a), the perfect polarization conversion is achieved if $\vec E^{r}$ can be expressed as a $ 90^{\circ} $-counterclockwise  rotation of $\vec E^{i}$. Consequently, $\vec E^{r}$ can be expressed as a function of the rotation matrix $ \underline{\underline{R}}\left( \theta \right) $ as follows:

\begin{equation}
{\vec E^{r}} = \underline{\underline{R}}\left( \theta = \frac{\pi}{2} \right) {\vec E^{i}}, 
\end{equation}
with:
\begin{equation}
 {\underline{\underline{R}}\left( \theta \right) } = \begin{bmatrix}
                    \cos \theta  &  - \sin \theta   \\[5pt]
                    \sin \theta  & \cos \theta    \\
                  				\end{bmatrix} \\				
\end{equation}

Therefore, the adopted strategy provides a perfect polarization conversion if the two components ($ x $ and $ y $) of the impinging electric field are reflected with the same amplitude and with a phase difference equal to $ 2\pi $: 

\begin{equation}
\begin{cases} E_x^{r} = -E_x^{i} \\[3pt] E_y^{r} = E_y^{i}  \end{cases}  \Rightarrow  \begin{cases} |\Gamma_x|=|\Gamma_y| \\[3pt]  \angle{\Gamma_x}=-\angle{\Gamma_y}  \end{cases}.
\label{eq:Pol_conditions}
\end{equation}

Indeed, both $ x $ and $ y $ components are completely reflected but with a different phase: the field component orthogonal to the dipole is subjected to a reflection coefficient equal to -1 whereas the electric field component parallel to the dipole is reflected with a reflection coefficient equal to +1.

Finally, the copolar ($ E^{co} $) and crosspolar ($ E^{cr} $) components of the reflected electric field can be expressed as follows:

\begin{equation}
E^{co}= E^{r}_x cos(\varphi)+ E^{r}_y sin(\varphi),
\end{equation}

\begin{equation}
E^{cr}=-E^{r}_y cos(\varphi)+E^{r}_x sin(\varphi).
\end{equation}

The polarization conversion performance of the dipole resonator as a function of the rotation angle $ \varphi $ are reported in Fig.~\ref{fig:PolConvDipole}(c). It is evident from the figure that the perfect polarization conversion is obtained when $ \varphi = 45^{\circ} $ that is the rotation for which the conditions reported in eq.~\ref{eq:Pol_conditions} are verified.

It is worth underlining that the polarization conversion mechanism is valid regardless of the particular unit cell topology provided that the reflection coefficients fullfill the aforementioned conditions.

\bibliographystyle{IEEEtran}
\bibliography{references} 

\end{document}